\documentclass[9pt,twocolumn,twoside]{optica}

\newcommand\rr{\mathbf{r}}
\newcommand\dd{\mathrm{d}}

\newcommand{\norm}[1]{\left\Vert #1\right\Vert}

\setboolean{shortarticle}{false}

\title{Photoacoustic fluctuation imaging: theory and application to blood flow imaging.}

\author[1]{Sergey Vilov}
\author[1]{Guillaume Godefroy}
\author[1]{Bastien Arnal}
\author[1,*]{Emmanuel Bossy}

\affil[1]{Univ. Grenoble Alpes, CNRS, LIPhy, 38000 Grenoble, France}
\affil[*]{Corresponding author: emmanuel.bossy@univ-grenoble-alpes.fr}

\begin{abstract}
Photoacoustic fluctuation imaging, which exploits randomness in photoacoustic generation, provides enhanced images in terms of resolution and visibility, as compared to conventional photoacoustic images. While a few experimental demonstrations of photoacoustic fluctuation imaging have been reported, it has to date not been described theoretically. In the first part of this work, we propose a theory relevant to fluctuations induced either by random illumination patterns or by random distributions of absorbing particles. The theoretical predictions are validated by Monte Carlo
finite-difference time-domain simulations of photoacoustic generation in random particle media. We provide a physical insight into why visibility artefacts are absent from second-order fluctuation images. In the second part, we demonstrate experimentally that harnessing randomness induced by the flow of red blood cells produce photoacoustic fluctuation images free of visibility artefacts. As a first proof of concept, we obtain two-dimensional images of blood vessel phantoms. Photoacoustic fluctuation imaging is finally applied in vivo to obtain 3D images of the vascularization in a chicken
embryo.
\end{abstract}

\setboolean{displaycopyright}{true}

\begin{document}

\maketitle

\section{Introduction}

Photoacoustic (PA) imaging is a widely-spread biomedical imaging modality that makes use of ultrasound waves resulting from light absorption~\cite{beard2011biomedical}. In acoustic-resolution PA imaging, arrays of ultrasound detectors are usually used to record PA signals generated by absorbing structures. PA reconstruction algorithms (such as delay-and-sum or backpropagation algorithms) are then used to provide images from a set of PA signals. PA images may, however, present artefacts that arise from coherence of PA waves and characteristics of the detection system such as geometry and frequency bandwidth. Limited-view artefacts occur when strongly anisotropic absorbing structures (such as blood vessels) coherently emit PA waves in some preferential directions, and these waves can not be detected due to the finite aperture and/or directivity of the ultrasound detectors. In such a case, some parts of the absorbing structure are not visible on the reconstructed PA image. Visibility artefacts also occur with resonant ultrasound detectors, which filter out low-frequency components of PA waves emitted by large absorbers (large as compared to the detection wavelength range). For instance, for large blood vessels, only the vessel boundaries may be visible in PA images.

Various solutions have been proposed to overcome the limited-view problem with finite-aperture detectors. One approach consists in enhancing the detection aperture by performing multiple PA acquisitions at different angles. This can be done by rotating the ultrasound probe around the object~\cite{yang2007fast} or by spinning the object itself~\cite{kruger2003thermoacoustic}. The detection aperture can also be augmented by placing acoustic reflectors  at edges of the imaging zone~\cite{huang2013improving,li2015tripling}. 
Alternatively, the imaged object and the ultrasound probe can be placed inside a reverberating cavity~\cite{cox2007photoacoustic, ellwood2014photoacoustic, cox2009photoacoustic}. This allows detecting PA waves emitted in all possible directions even when only a single-element probe is used~\cite{cox2009photoacoustic}. 

All the mentioned techniques require that the whole range of imaging angles is accessible for detection. However, this is not necessarily feasible in real clinical environment. In addition,  mechanical scanning of the detectors or sample increases the acquisition time and, therefore, degrades the temporal resolution. 
Methods that do not require detecting waves emitted in all directions have also been proposed. Sub-wavelength sparsely distributed absorbing particles act individually as isotropic PA sources. This can been exploited to remove limited-view artefacts: Dean-Ben \textit{et al.} demonstrated  that visibility  in the limited-view scenario can be improved by means of  the localization approach~\cite{dean2018localization}, or by using a nonlinear combination of tomographic reconstructions representing sparsely distributed moving particles~\cite{Dean-Ben:17}. Unless \emph{sparsely} distributed contrast agents are used~\cite{zhang2019vivo}, these techniques remain limited in practice as blood consists of a \emph{concentrated} suspension of red blood cells. 
One more approach consists in heating tissue locally with a focused ultrasound beam and thus generating artificial PA sources via the temperature dependence of the Gruneisen coefficient~\cite{Wang:15}. By scanning the focused ultrasound beam across the sample and accumulating the resulting PA images, the whole object is reconstructed. However, this approach is limited in practice by the safety thresholds of focused ultrasound. Similar to all other aforementioned approaches this method requires a very long acquisition time leading to a low temporal resolution. 
Recently, deep-learning based approaches have also been proposed as a way to palliate visibility artefacts from limited-view detection in photoacoustic imaging~\cite{kim2020deep, davoudi2019deep,godefroy2020solving}. While powerful, deep-learning based imaging methods also have limitations. These include the availability of a training set and the performances inherently limited to unseen samples "close" enough to those of the training set, which is a severe practical limitation.

An approach based on multiple-speckle illumination was proposed by Gateau \textit{et al.}~\cite{gateau2013improving}.  In this experimental work, a random intensity distribution of speckle patterns, which changed from pulse to pulse, induced fluctuations in each pixel of the corresponding series of PA images. It was demonstrated experimentally with free-space-generated series of speckle patterns that a second-order fluctuation image provided a faithful representation of the absorbing distribution, free of limited-view and limited-bandwidth artefacts. However, exploiting optical speckle illumination for imaging tissue at depth turns out to be  challenging since the fluctuation signal is expected to be very small when the speckle grain size is of the order of the optical wavelength. As a consequence, this technique has so far never been demonstrated in a close-to-clinical environment. 
Moreover, the work by Gateau \textit{et al.}~\cite{gateau2013improving} provided no clear theoretical explanation for the visibility enhancement in fluctuation imaging with multiple speckle illumination. Since that  work, there have been other studies based on PA signals fluctuations. In particular, such fluctuations were deployed to obtain super-resolution in PA imaging~\cite{chaigne2016super,hojman2017photoacoustic,murray2017super,chaigne2017super}. Most generally,  PA fluctuation imaging utilizes some randomness in PA generation to provide enhanced images as compared to conventional PA imaging.  While it has been successfully demonstrated experimentally for super-resolution imaging (induced either from multiple-speckle illumination~\cite{chaigne2016super,hojman2017photoacoustic,murray2017super} or from random distributions of absorbing particles~\cite{chaigne2017super}) and for visibility enhancement (with multiple-speckle illumination~\cite{gateau2013improving}), there is, however, no comprehensive theoretical description of fluctuation PA imaging to date. 

In the first part of this work, we propose a unified theoretical framework relevant to both fluctuations induced by random illumination patterns and fluctuations induced by random distributions of absorbing particles. The theoretical considerations are validated in Monte-Carlo finite-difference time-domain (FDTD) simulations of PA generation in a random medium. In particular, we explain why visibility artefacts are absent in $2^{nd}$-order fluctuation images. This property turns out to be completely independent of the origin of fluctuations. In addition,   our theoretical results provide a quantitative comparison between fluctuations from multiple-speckle illumination and fluctuations from random distributions of absorbing particles.
In the second part, we demonstrate experimentally that fluctuations induced by a blood flow can be exploited under physiological conditions to produce PA fluctuation images free of visibility artefacts, as opposed to  conventional PA imaging. Two-dimensional images of vessel-mimicking phantoms flown with blood at physiological concentrations are obtained as a first  proof-of-concept demonstration. Finally, the approach is validated  \textit{in vivo} by performing 3D imaging of a chicken embryo.  

\section{Theory}

In this section, we propose a unified theoretical framework for PA fluctuation imaging applicable to both fluctuations induced by random illumination patterns and fluctuations induced by random distributions of absorbing particles. We first provide theoretical expressions for the mean PA image and the fluctuation PA image. These expressions are then used to explain why the fluctuation images do not possess the visibility artefacts observed on the mean images (equivalent to conventional PA images). Finally, we discuss the dependence of the amplitude of the fluctuation image on characteristic statistical properties (mean, variance, characteristic size) of the random process inducing fluctuations (illumination patterns or distribution of absorbing particles). In particular, we compare the case of randomly-generated multiple-speckle illumination to that of randomly-distributed red blood cells, and discuss quantitatively why flowing red blood cells lead to PA fluctuations much larger than those produced by multiple speckle illumination deep in tissue.

\subsection{Analytical predictions}

\subsubsection{Theoretical framework}

PA imaging provides images of the absorbed energy density $\alpha$ that we describe in this work as:
\begin{equation}
    \alpha_k(\rr)=
    \mu_0 F_0 [g_k(\rr) \times f(\rr)]
    \label{eq:alpha}
\end{equation}
In the expression above, $F_0$ has the dimension of a light fluence, $\mu_0$ has the dimension of an  absorption coefficient; $f(\rr)$ is a dimensionless binary function that describes the structure containing optical absorbers (for instance  blood vasculature with red blood cells) : $f(\rr)=1$ inside the imaged structure, $f(\rr)=0$) outside; $g_k(\rr)$ is the $k^{th}$ realization of the random process inducing fluctuations: it may either describe a spatial distribution of light intensity or a spatial distribution of absorbers. For random illumination patterns, such as speckle patterns~\cite{gateau2013improving}, $F_0\times g_k(\rr)$ (with $g_k$ being a continuous real-valued function) is the associated  random fluence corresponding to each laser shot $k$, and the absorption coefficient $\mu_0$ is considered homogenous in the absorbing structure $f(\rr)$. For random distributions of absorbing particles, such as red blood cells, $\mu_0\times g_k(\rr)$ is the associated heterogeneous absorption coefficient for each laser shot $k$ while $g_k(\rr)$ is a binary function equal to 1 inside the particles, and 0 elsewhere. 

From the most general perspective, $g_k(\rr) \times f(\rr)$ thus describes the spatial distribution of absorbed energy. We assume that $g_k(\rr)$ is a spatially stationary and isotropic random process. Its relevant statistical properties in the context of our work are the mean $\eta=<g_k(\rr)>_k$ and  the variance $\sigma_g^2=<(g_k(\rr)-\eta)^2>_k$ as first-order statistical properties and its autocovariance $$C(\rr_1,\rr_2)=C(\norm{\rr_1-\rr_2})=<(g_k(\rr_1)-\eta)(g_k(\rr_2)-\eta)>_k$$ as a second-order statistical property. We note that $C(\boldsymbol{0})=\sigma_g^2$. We further assume that the normalized autocovariance $C(\norm{\rr_1-\rr_2})/C(\boldsymbol{0})$ is a function with a finite volume $V_g$ and characteristic width $D_g$ (with $D_g^n=V_g$, n=2 or n=3 depending on the relevant dimensionality):
\begin{equation}
\label{eq:DefinitonDg}
\int_{\mathbb{R}^n} C(\norm{\rr})d\rr=\sigma_g^2V_g=\sigma_g^2D_g^n 
 \end{equation}

For multiple-speckle illumination, $V_g$ (resp. $D_g$) is of the order of the characteristic volume (resp. size) of a speckle grain. For random distributions of mono-dispersed absorbing particles, $V_g$ (resp. $D_g$) is of the order of the particle characteristic volume (resp. size). 

We assume that each PA reconstruction $A_k(\rr)$ corresponding to laser shot $k$ can be expressed as the convolution between $\alpha_k(\rr)$ and the point spread function (PSF) $h(\rr)$ of the imaging system:
\begin{equation}
    A_k(\rr)= \Gamma \cdot \alpha_k(\rr)\ast h(\rr),
    \label{eq:model}
\end{equation}
where $\Gamma$ is the Grüneisen parameter. The convolution in Eq.~\ref{eq:model} implies a space-invariant PSF, which is a valid assumption within the small fields of view that we deal with in this work. We consider the most general framework where the PSF $h(\rr)$ is either a real-valued function or a complex-valued function. A real-valued PSF corresponds to reconstruction based on real-valued RF signals, whereas a complex-valued PSF corresponds to reconstruction based on complex-valued signals. 

Complex-valued reconstruction provides a straightforward way to remove radio-frequency oscillations by using the modulus of the complex-valued reconstruction as the final image, and is a widely used approach in medical ultrasound imaging. In the context of photoacoustic imaging, an illustration of the differences between real-valued only images and complex-modulus images may be found in a previous publication from our group~\cite{chaigne2017super}.

\subsubsection{Mean photoacoustic image}

We consider the mean PA image defined by $E[A](\rr)=<A_k(\rr)>_k$, where $<.>_k$ denotes an ensemble average (average over laser shots in practice). As $\eta=<g_k(\rr)>_k$, $E[A](\rr)$ can be straightforwardly expressed as:
\begin{equation}
    E[A](\rr) =\Gamma F_0\mu_0 \eta [f(\rr)\ast h(\rr)]
    \label{eq:mean_image}
\end{equation}
In the case of random distributions of absorbing particles, $\eta=<g_k(\rr)>_k$ corresponds to the volume fraction of absorbers, and $\eta\mu_0$ is the average absorption coefficient of the imaged structure. In the case of random illumination patterns and a homogeneously absorbing structure, $<g_k(\rr)F_0>_k=\eta F_0$ corresponds to the average light fluence inside the imaged structure, and $\mu_0$ is the absorption coefficient of the homogeneously absorbing imaged structure.  Most generally, Eq. (\ref{eq:mean_image}) corresponds to conventional PA reconstruction of a homogeneously absorbing structure described by $f(\rr)$. Eq. (\ref{eq:mean_image}) is independent of the real-valued or complex-valued nature of $h(\rr)$. For real-valued reconstruction, $E[A](\rr)$ directly represents the mean PA image, whereas for complex-valued reconstruction, the mean PA image is rather represented by the modulus $\left | E[A](\rr)\right |$. 

\subsubsection{Photoacoustic fluctuation image}

In the context of this work, we define the PA fluctuation image as a second-order fluctuation image, defined as the square root of the variance image $\sigma^2[A](\rr)=<[ A_k(\rr)-E[A](\rr)]^2>_k=<[A_k(\rr)-E[A](\rr)]\times[A_k(\rr)-E[A](\rr)]^*>_k$, for both  real-valued and complex-valued reconstructions. Higher-order PA fluctuation images may also be defined~\cite{chaigne2017super}, but are beyond the scope of this work. The variance  image $\sigma^2[A](\rr)$ can be expressed by use of the autocovariance function $C(\rr_1,\rr_2)$:
\begin{equation*}
\begin{split}
    &\sigma^2[A](\rr)=\Gamma^2 \mu_0^2 F_0^2<\int_{\rr_1}(g_k(\rr_1)-\eta)f(\rr_1)h(\rr-\rr_1)\dd \rr_1\times\\
    &\int_{\rr_2}(g_k(\rr_2)-\eta)f(\rr_2)h^*(\rr-\rr_2)\dd \rr_2>_k=\\
    &=\Gamma^2\mu_0^2F_0^2\iint C(\rr_1,\rr_2)f(\rr_1)f(\rr_2)h(\rr-\rr_1)h^*(\rr-\rr_2)d\rr_1d\rr_2
    \label{eq:conv_variance_strict}
\end{split}
\end{equation*}
As the fundamental and major assumption of our work, we now assume that the characteristic size $D_g$ of the random distribution is much smaller than the shortest characteristic sizes of both $h(\rr)$ and $f(\rr)$. In the context of random distributions of red blood cells, $D_g$ is of the order of 10 $\mu$m.  In the context of speckle illumination, $D_g$ is the typical size of a speckle grain and is therefore not larger than 1  $\mu$m for visible or near-infrared light.  Moreover, the shortest characteristic length scale of $h(\rr)$ for resonant transducers is the wavelength $\lambda$ at the central frequency. Our assumption is therefore fully verified for vessels at least several tens of  $\mu$m in diameter and central frequencies up to several tens of MHz (acoustic wavelengths being not smaller than several tens of  $\mu$m), which corresponds to the most of practical situations of interest. 
Under this assumption,  $C(\rr_1,\rr_2)$ may be replaced in the integral by $\left [ \int_{\mathbb{R}^p} C(\norm{\rr'})d\rr' \right ]\times \delta(\rr_1-\rr_2)=\sigma^2_g V_g \delta(\rr_1-\rr_2)$:
\begin{equation*}
\begin{split}
\sigma^2[A](\rr) &= \Gamma^2\mu_0^2F_0^2\times\\
&\iint \sigma^2_g V_g \delta(\rr_1-\rr_2)f(\rr')f(\rr_2)h(\rr-\rr_1)h^*(\rr-\rr_2)\dd \rr_1\dd\rr_2
\end{split}
\end{equation*}
This leads to the following final expression for the PA fluctuation image, valid for both real-valued and complex-valued PSF:
\begin{equation}
    \sigma[A](\rr) =  \Gamma \mu_0 F_0 \sigma_g \sqrt{V_g}\sqrt{f^2(\rr) \ast |h|^2(\rr)}
\label{eq:fluctuation_image}
\end{equation}
While it has already been shown by our group that the PA fluctuation image could be written as a convolution involving the square of the PSF~\cite{chaigne2016super,chaigne2017super}, Eq. (\ref{eq:fluctuation_image}) for the first time explicitly links  the amplitude of the fluctuation image to statistical properties of the random process at stake (through its mean $\eta$, its standard deviation $\sigma^2_g$ and the volume of its normalized autocovariance $V_g$). In particular, Eq. (\ref{eq:fluctuation_image})  remains valid when the characteristic size of $f(\rr)$ is smaller than that of $h(\rr)$, which is of interest for super-resolution imaging: as opposed to our previous works on fluctuation-based super-resolution imaging~\cite{chaigne2016super,chaigne2017super}, we here provide a quantitative  prediction of the second-order super-resolution photoacoustic fluctuation image, which is relevant for quantitative analysis of such images. In this work, we however focus on the consequences of Eq. (\ref{eq:fluctuation_image}) on the visibility problem.

\subsection{Consequences on visibility artefacts}

Eq. (\ref{eq:fluctuation_image} ) indicates that the fluctuation image involves a convolution with the square of the PSF. This is a well-known result from the Super-resolution Optical Fluctuation Imaging (SOFI) approach initially proposed for super-resolution imaging relying on blinking fluorophores. The SOFI approach permits obtaining super-resolved images based on the fact that the PSF to the $2^{nd}$ power is sharper than the PSF itself \cite{dertinger2009fast}. Our group adapted the SOFI approach to super-resolution PA imaging, with fluctuations induced from either multiple-speckle illumination~\cite{chaigne2016super} or moving absorbers~\cite{chaigne2017super}. Although this convolution with the squared PSF resulting in the $2^{nd}-$order fluctuation image has been known both for optical and PA imaging, its effect was analyzed only from the resolution perspective~\cite{chaigne2016super,chaigne2017super}. However, as we will now discuss, it also has a direct impact in the context of both the limited-view and resonant-transducer problems.

\subsubsection{Illustration with simulation results}
\label{subsubsection:Illustration}
\begin{figure}[!h]
\centering
\includegraphics[width=\linewidth]{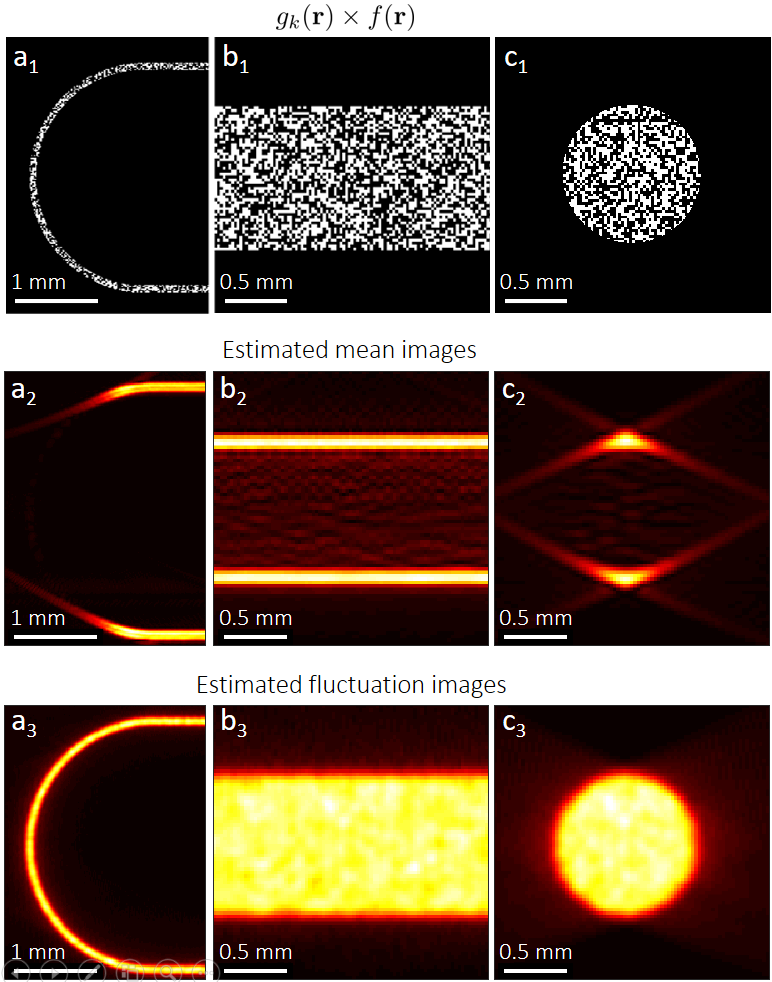}
    \caption{Two-dimensional simulation results from complex valued images. $a_1-c_1$: typical realization for each of the three different structures $f(\rr)$ with $\eta=50\ \%$. For all three objects, the probe consists of a linear array (N = 128 elements, 15 MHz central frequency, bandwidth $80\%$, pitch 110 $\mu$m) located 15 mm away above the objects. The corresponding experimental configurations are shown in Fig.\ref{fig:configurations}. $a_2-c_2$: modulus of the mean photoacoustic reconstruction obtained for each type of structure. $a_3-c_3$: Corresponding fluctuation images, free of visibility artefacts. Mean and fluctuation images are estimated from  N = 100 realizations.} 
\label{fig:SimulationResults}. 
\end{figure}

To illustrate  our discussion, we first use two-dimensional (2D) numerical simulations  showing how fluctuation imaging removes visibility artefacts. 
The numerical simulations were carried out with the 2D version of a finite-difference time-domain (FDTD) freely-available code developed by our group~\cite{Simsonic}. The simulations output consisted of a set of RF signals detected by a linear transducer array similar to that used in the experiments (central frequency 15 MHz, bandwidth $80\%$, N = 128 elements, pitch 110 $\mu$m). The input of each simulation consisted of a source image defined on a cartesian grid with a spatial step of 2.5 $\mu$m.

Three different structures $f(\rr)$ were used to illustrate how  PA fluctuation imaging removes  artefacts of conventional PA imaging, as illustrated in Fig.~\ref{fig:SimulationResults} (top row). Similar structures used in the phantoms experiments are represented in Fig.\ref{fig:configurations}. 
The simulation input for each realization of the random process consisted of a random source image $g_k(\rr)\times f(\rr)$. The distribution $g_k(\rr)$ was represented by an image of super-pixels (group of $q\times q$ pixels on the simulation grid), that was randomly assigned binary values 0 or 1 following a Bernouilli distribution with mean value $\eta$. The simulations permitted producing the full range of $\eta=<g_k>$ = 0 - 100\%. In this model, $\eta$ represents the surface fraction of the absorbed light and can be varied from 0 to $100\ \%$. 

Fig.\ref{fig:SimulationResults} shows some example results obtained in our Monte-Carlo simulations. To obtain these results, the surface fraction of absorbed energy was set to $\eta=50\ \%$ and the size of each individual light absorbing patch was $20 \ \mu m \times 20\ \mu m$ ($16\times 16$ super-pixel). Fig.\ref{fig:SimulationResults} shows the modulus images obtained from complex-valued signals. 
Fig.\ref{fig:SimulationResults} (column a) illustrates the limited-view problem: conventional PA imaging (middle row) only shows features parallel to the probe, while PA fluctuation imaging (bottom row) renders the full structure correctly. Fig.\ref{fig:SimulationResults} (column b) illustrates the limited-bandwidth problem: conventional PA imaging (middle row) only shows the upper and lower boundaries (high spatial frequencies) of the imaged object and does not reveal its inner content (low spatial frequencies), while PA fluctuation imaging (bottom row) correctly renders the full structure. Fig.\ref{fig:SimulationResults} (column c) illustrates both  problems  and confirms the outperformance of fluctuation imaging (bottom row).  Such outperformance have first been demonstrated experimentally in our group several years ago in the specific context of multiple-speckle illumination~\cite{gateau2013improving}, but without any clear theoretical explanation. 

To illustrate the principle of the proposed approached,  the simulations results presented in Fig.1 were obtained without any additional sources of fluctuations (such as detection noise or laser pulse energy fluctuations). Additional simulation results presented in the Supplementary Information (section 4)  confirms that our approach also works very efficiently under noisy environment:  pulse laser energy fluctuations may be compensated for via SVD filtering; the fluctuation of interest may be detected on top of a constant noise background provided that a sufficiently large number N of measurements are used to estimate the fluctuations image. Our simulation results with noise show that N may range in practice from a few units to a few hundred (see Fig. S3 of the Supplementary Information), depending on the ratio between the fluctuation of interest and the noise level (FNR, see section 3 of the Supplementary Information.

\subsubsection{Mechanism of visibility enhancement in fluctuation imaging}

\begin{figure}[t]
\centering
\includegraphics[width=1.0\linewidth]{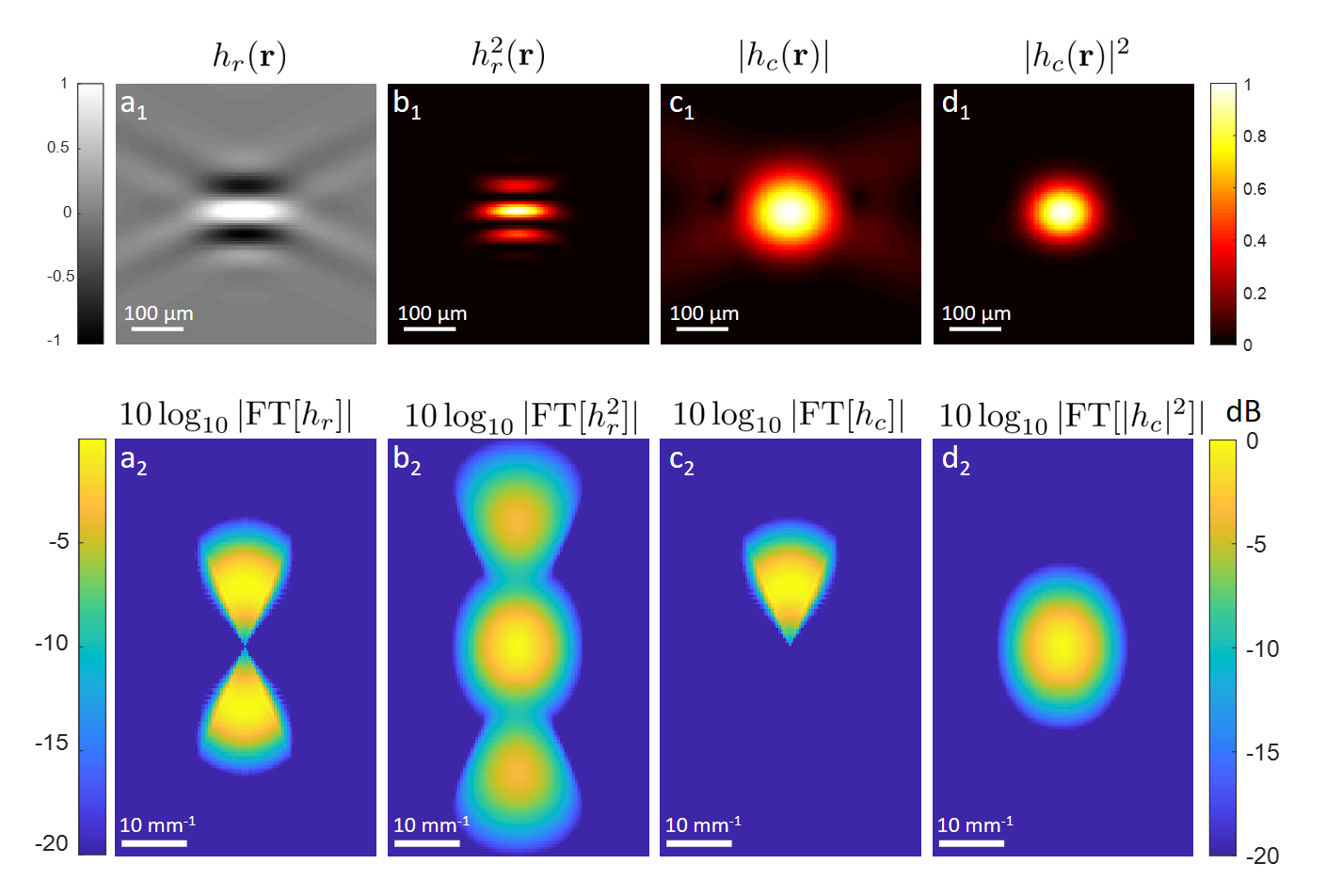}
\caption{Two-dimensional simulation results: typical PSF $h(\rr)$ and $h^2(\rr)$ for limited-view photoacoustic imaging with resonant transducer, for both the real-valued PSF ($h_r(\rr)$)  and the complex-valued PSF ($h_c(\rr)$). Top row: spatial domain. Bottom row: Fourier domain. All functions are normalized to one. The PSF was derived from FDTD-computed signals, with the same 1D linear array as that used for Fig. 1, with a point source 15 mm away from the transducer.}
\label{fig:PSF}
\end{figure}

Based on the expression of the fluctuation image involving a convolution with the squared PSF, we now explain why PA fluctuation imaging removes visibility artefacts. 
The PSF corresponding to the images shown in Fig.\ref{fig:SimulationResults} was computed in the same  simulation setting by placing a point source in the center of the field of view (FOV) at the distance of $z=15$ mm from the probe. We note $h_r(\rr)$ the real-valued PSF and $h_c(\rr)$ its complex-valued counterpart for the considered 1D linear transducer array (aligned along the horizontal direction as in  Fig.\ref{fig:configurations}). Fig.\ref{fig:PSF} shows the PSF $h(\rr)$ and its square modulus $|h|^2(\rr)$  along with the corresponding spatial Fourier transforms. 

The bipolar nature of $h_r(\rr)$ and $h_c(\rr)$ for resonant transducers leads to low spatial frequency components of the objects being filtered out, which can be explained with Fig.\ref{fig:PSF}.($a_2$) and  Fig.\ref{fig:PSF}.($c_2$) showing the PSF in the Fourier space. 
Moreover, the anisotropy of $h_r(\rr)$ and $h_c(\rr)$ leads to a selective detection of only spatial frequencies within the numerical aperture of the transducer. Previous analyses of $|h|^2(\rr)$  in the context of super-resolution PA imaging used its extended support in the Fourier space as compared to that of $h(\rr)$, to explain the obtained super-resolution (see Supplementary Information in ~\cite
{chaigne2016super,chaigne2017super}). However, it also follows from the support of $h_r^2(\rr)$ and $|h_c|^2(\rr)$ in the Fourier space (see Figs.\ref{fig:PSF}.($b_2$) and  Fig.\ref{fig:PSF}.($d_2$)) that imaging with the squared modulus of the PSF preserves low spatial frequencies in all directions, as opposed to $h(\rr)$.  As a consequence, the fluctuation image does not suffer from the limited-view and resonant-bandwidth artefacts observed in conventional imaging. 

Fig.\ref{fig:PSF}.($c_1$) and ($d_1$) also illustrate how using the modulus of the complex valued PSF eliminates the oscillatory behaviour of $h_r(\rr)$ (Fig.\ref{fig:PSF}.($a_1$) and ($b_1$)). 
Interestingly, the conclusions above drawn from the analysis in the Fourier space may also be obtained in the physical space by considering the difference between coherent and incoherent summation. Since $h(\rr)$ is a bipolar function, its convolution with absorbing objects which are always positive may lead to destructive interference, which in practice occurs when dealing with large objects or those that emit PA waves away from the detectors. As opposed to $h(\rr)$, $h^2(\rr)$ is positive only and can therefore not lead to destructive interference. In other words, conventional imaging involves coherent summation of bipolar PA waves from positive sources, that may lead to destructive interference, whereas fluctuation imaging involves  summation of positive fluctuations that can never interfere destructively. One can draw a parallel with optics where amplitudes sum up coherently while intensities sum up incoherently.       

\subsection{Amplitude of the fluctuation image}

The mechanism discussed and illustrated above with Monte-Carlo simulations and a simple random model is independent of the nature of fluctuations. These fluctuations may arise either from random heterogeneous illumination patterns (such as speckle patterns) or from random distributions of optical absorbers (such as red blood cells). However, the amplitude of the fluctuation image fundamentally depends on the statistical properties of the relevant random process (mean $\eta$,  standard deviation $\sigma^2_g$, typical length scale $D_g$), as indicated by Eq.\ref{eq:fluctuation_image}. To illustrate our discussion, we consider the fluctuation amplitude expected from structures much larger than the PSF. In this case, $\sqrt{f^2(\rr) \ast |h|^2(\rr)}\simeq \sqrt{\int_{\rr'} |h|^2(\rr')\dd \rr'}=||h||_2$, and the fluctuation amplitude is given by: \begin{equation}
    \sigma[A] =  \Gamma \mu_0 F_0 \sigma_g \sqrt{V_g}||h||_2
\label{eq:fluctuation_INSIDE}
\end{equation}  
while the average PA value is expected to be zero ($E[A] \propto f(\rr) \ast h(\rr)\propto \sqrt{\int_{\rr'} h(\rr')\dd \rr'}=0$), as a consequence of the resonant bandwidth artefact (bipolar nature of $h(\rr)$). In practice, PA fluctuation amplitude must be larger than other fluctuations such as the the detection noise to be measurable. The fluctuation amplitude is in part determined by the sensitivity of the system (quantified through $||h||_2$), and by the statistical properties of the random distribution of absorbed energy ($\eta$, $\sigma_g$ and $V_g$).

We first demonstrate by use of 2D numerical simulations that the fluctuation amplitude depends on $\eta$ and $V_g=D_g^2$ as predicted by our theory. To this end, we shall use the simple random model already introduced in section \ref{subsubsection:Illustration}. Then we provide more specific expressions for the fluctuation amplitude in the case of multiple speckle illumination and in the case of random distributions of absorbing particles. We finally discuss why fluctuations from absorbing particles such as red blood cells are expected to be much larger than those from multiple speckle illumination deep in tissue. We recall that equations \ref{eq:fluctuation_image} and \ref{eq:fluctuation_INSIDE} are  valid only for characteristic lengths $D_g$ much smaller than the shortest dimension of the PSF (about $50\ \mu \mathrm{m}$ as half a wavelength in the context of our work, see Fig.~\ref{fig:PSF}).

\subsubsection{Validation with simulations}

To validate our theory with simulations, we consider a simple model of a random distribution of absorbed energy for which $V_g$ (and thus $D_g$) is independent of $\eta$. To do so, we use the model already introduced  in section~\ref{subsubsection:Illustration}, for which $g_k$ is defined on a cartesian grid with pixel/voxel size $D$. Each pixel of the grid may acquire a random binary value 0 or 1 from a Bernoulli distribution with  mean value of $\eta$. We recall that $\eta$ defined as such corresponds to the average surface/volume fraction $\eta=<g_k(\rr)>_k$ of the corresponding random medium. The variance for the Bernoulli distribution is given by $\sigma_g^2=\eta\times (1-\eta)$.
Moreover, it can be shown straightforwardly for this random medium that $C(\rr',\rr'')=\eta\times (1-\eta)\prod_{i=1}^n\Lambda_D(x_i'-x_i'')$, with n being the space dimension (n=2 or n=3) and $\Lambda_D$ being the 1D unit triangular function with support $2\times D$. From Eq.\ref{eq:DefinitonDg}, one obtains that $V_g=D^n$ (i.e. $D_g=D$), independently of $\eta$. The expression for the fluctuation amplitude for this random distribution of absorbed energy becomes :
\begin{equation}
    \sigma[A](\rr) =  \Gamma \mu_0 F_0 \sqrt{\eta(1-\eta)D^n}||h||_2
\label{eq:BernoulliPixel}
\end{equation}

\begin{figure}[!t]
\centering
\includegraphics[width=\linewidth]{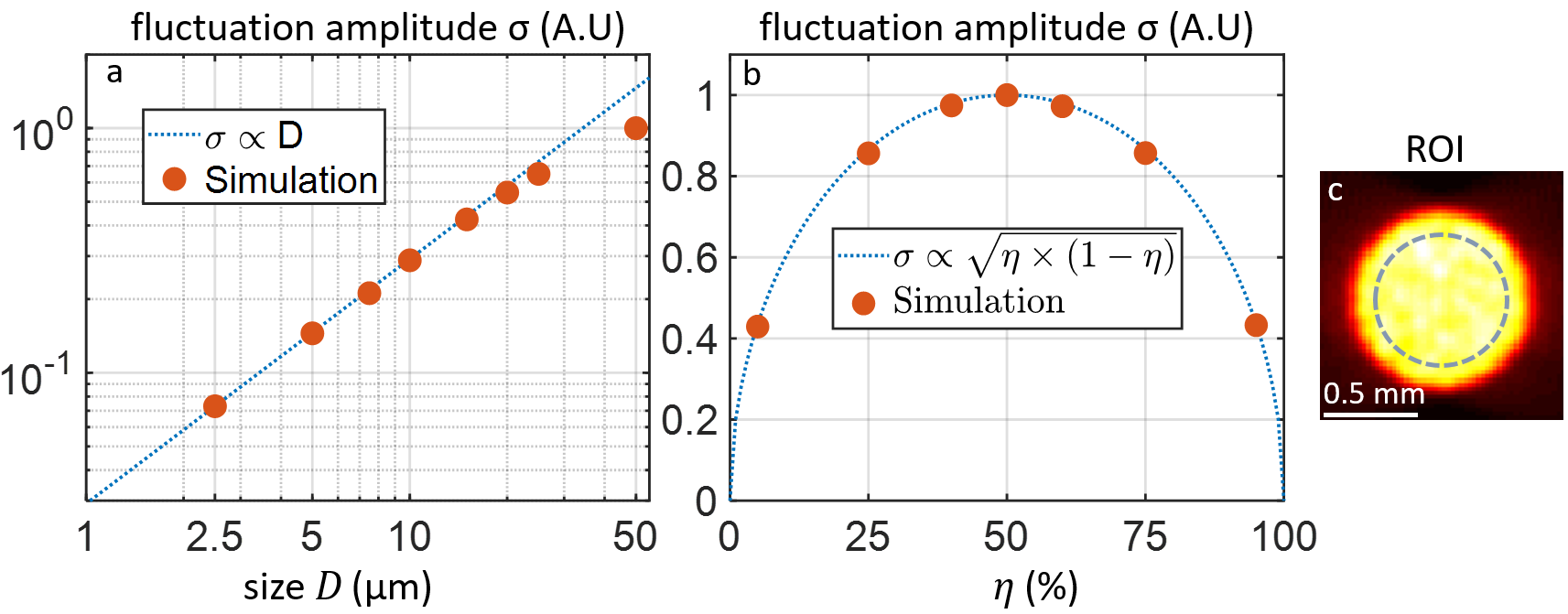}
\caption{ Two-dimensional FDTD simulation results. The fluctuation amplitude $\sigma[A]$ was estimated over a circular region of interest (ROI) for the disk configuration. The absorbed energy is distributed over  square pixels of size $D$, with probability $\eta$. (a) Dependence of $\sigma[A]$ as a function of $D$ (for a fixed value $\eta=50\ \%$). (b) Dependence of $\sigma[A]$ as a function of $\eta$ (fixed size $D=20\ \mu m$). The theoretical dependence are those given by Eq.\ref{eq:BernoulliPixel} with $n=2$. }
\label{fig:TheoryVsSimulations}
\end{figure}

To validate the predictions of Eq.\ref{eq:BernoulliPixel}, we ran 2D simulations ($n=2$) for the disk structure (much larger than the PSF) with different values of $\eta$ and $D$. The fluctuation amplitude was estimated by spatially averaging the fluctuation image over a circular region of interest (see ROI in Fig.\ref{fig:TheoryVsSimulations}) . As illustrated in Fig.\ref{fig:TheoryVsSimulations}, the dependence of the fluctuation amplitude on $\eta$ (for fixed $D=10\ \mu m$) and on the pixel size  $D$ (for fixed $\eta=50\ \%$) measured from the  FDTD simulations ($N=100$ random realizations for each set of parameters) exactly matches the one following from Eq.\ref{eq:BernoulliPixel}. In particular, the dependence on $\eta$ follows $\sqrt{\eta\times(1-\eta)}$ and $\eta=50\ \%$ is the value that maximizes the fluctuation amplitude. As expected, the size dependence diverges from the theory when $D$ approaches half a wavelength ($50\ \mu \mathrm{m}$). 

\subsubsection{Multiple speckle illumination}
We consider here a homogeneously absorbing structure $f(\rr)$ (constant absorption coefficient $\mu_0$), illuminated by random multiple speckle illumination described by the fluence $F_k(\rr)=F_0\times g_k(\rr)$. We further assume that the speckle is fully developed,  so the statistical properties of the fluence distribution are well-known: the intensity/fluence follows an exponential distribution, with $\sigma_F=<F>$, where $<F>=\eta F_0$ is the average intensity/fluence.  $D_g=D_s$ is a measure of the speckle grain size, and is independent of the average fluence $\eta F_0$. $V_s=D_s^n$ is the speckle grain area (i.e. $n=2$) or volume (i.e. $n=3$) depending on the relevant imaging configuration. The fluctuation amplitude in this case is thus given by:

\begin{equation}
    \sigma[A]_{\mathrm{speckle}} = \Gamma \mu_0 <F> \sqrt{V_s}||h||_2
\label{eq:FluctuationSpeckle}
\end{equation}

While it has long been acknowledged that detecting fluctuations from speckle illumination required large enough speckle grains, Eq.\ref{eq:FluctuationSpeckle} provides for the first time a quantitative theoretical prediction of the expected fluctuation amplitude as a function of all relevant parameters, including the geometry of the imaged object and the PSF of the imaging system. For a fixed average fluence, Eq.\ref{eq:FluctuationSpeckle} shows that the fluctuation amplitude is proportional to the square root of the area/volume of the speckle grain (depending on the relevant dimensionality).

\subsubsection{Random distribution of absorbing particles}

We now consider randomness that arises from the sample rather than from the illumination. In this case, the fluence is constant ($F_0$), and the absorption distribution is described by $\mu_k(\rr)=\mu_0\times g_k(\rr)$. More specifically, we consider randomness induced by random distributions of absorbing particles of a given characteristic volume $V_{p}$ (characteristic size $D_p=V_p^{1/n}$) inside the structure to image $f(\rr)$, such as red blood cells flowing in blood vessels. In this case, $g_k(\rr)$ is a binary function ($g_k(\rr)=1$ inside absorbing particles, $g_k(\rr)=0$ outside) and $\eta=<g_k(\rr)>_k$ simply represents the average volume fraction of absorbers. For blood, $\mu_0$ is  the absorption coefficient of pure hemoglobin (i.e. of a single red blood cell), and $\eta\mu_0$ is the average absorption coefficient of the whole blood as a suspension of red blood cells.  For $g_k(\rr)$ which is a binary random variable with $\eta=<g_k(\rr)>_k$ (Bernoulli distribution), the variance is $\sigma_g^2=\eta\times (1-\eta)$, regardless of the spatial structure of the random medium.  In this case, the expression for the fluctuation image becomes:
\begin{equation}
    \sigma[A]_{\mathrm{particles}} =  \Gamma \mu_0 F_0 \sqrt{\eta(1-\eta)V_g(\eta)}||h||_2
    \label{eq:FluctuationParticles}
\end{equation}
The characteristic volume $V_g(\eta)$, defined by Eq.\ref{eq:DefinitonDg}, is of the order of $V_p$, but its exact value generally depends on the volume fraction $\eta$, especially at high volume fraction where the positions of neighbors may become correlated. The dependence of $\sigma[A](\rr)$ to the volume of the absorber (for a fixed $\eta$) is exactly the same as the dependence to the size of the speckle grain:  the larger the particles, the larger the amplitude of the fluctuation image, within the limit that the particle size should be small compared to the PSF. For a given particle size and shape, the value of $\eta$ that maximizes the fluctuation amplitude is given by the maximum of $\eta(1-\eta)V_g(\eta)$, whose exact form depends on $V_g(\eta)$. In the simplest case where $V_g$ is independent of $\eta$, our theory predicts that the fluctuation amplitude is proportional to $\sqrt{\eta\times(1-\eta)}$, which is maximized for $\eta=50\ \%$. It is out of the scope of this work to investigate  the dependance of $V_g(\eta)$ for various types of media. Nonetheless, we demonstrate in the Supplementary information that $V_g(\eta)$ may be expressed as a function of $V_p$ through the so-called packing factor:
\begin{equation}
    V_g(\eta) = \frac{W(\eta)}{1-\eta}V_{p} 
\end{equation}
The packing factor $W(\eta)$ has been used in the field of ultrasound imaging to study the non linear dependence of echogeneicity as a function of $\eta$ in particles media ~\cite{twersky1978acoustic,shung1982ultrasound,twersky1987low,bascom1995fractal}, including blood~\cite{shung1982ultrasound,twersky1991,bascom1995fractal,savery2001point}. By use of the expression above, the PA fluctuation may thus also be expressed as
\begin{equation}
\sigma[A]_{\mathrm{particles}}=\Gamma\mu_0 F_0 \sqrt{\eta W(\eta) V_{p}}||h||_2
\end{equation}
which interestingly has the same dependence on $\eta\times W(\eta)$ as that reported for the backscattering coefficient~\cite{bascom1995fractal}, and agrees with recent experimental results~\cite{bucking2019acoustic} (see Supplementary Inf.)

\subsubsection{Multiple speckle illumination versus particles fluctuations}

As pointed out earlier, it is the ratio between the amplitude of the fluctuations of interest and the amplitude of other fluctuations (laser pulse energy  fluctuations, thermal noise) that determines the ability to detect the relevant fluctuations experimentally.  The theoretical expressions of the fluctuation images obtained for the case of multiple speckle illumination and random distributions of absorbers provide a quantitative comparison between the two situations. From Eqs. \ref{eq:FluctuationSpeckle} and \ref{eq:FluctuationParticles}, one gets:

\begin{equation}
\frac{\sigma[A]_{\mathrm{particles}}(\rr)}{\sigma[A]_{\mathrm{speckle}}(\rr)}=
\sqrt{\frac{\eta W(\eta)V_p }{V_s}}=
\sqrt{\eta W(\eta)\left(\frac{D_p}{D_s}\right)^n},
\label{eq:ComparisonSpeckleParticles}
\end{equation}
where it is assumed that the average fluence and the absorption coefficient are identical in both cases.  Applied to red blood cells ($V_p\sim 100\ \mu m^3$, $\eta\sim 50\ \%$, $W(50\%)\sim 16\ \%$~\cite{bascom1995fractal} and Supp. Inf) and and speckle grains in the multiple scattering regime deep in tissue ($D_s\sim \frac{\lambda}{2}\sim 0.25\ \mu m$ in the visible range), the relation above  (with $n=3$) indicates that fluctuations arising from red blood cells as absorbing particles are at least one order of magnitude   larger than fluctuations expected from multiple speckle illumination. 

\subsection{Conclusions from the theory}

In this part, we provided a unified theoretical framework to predict the amplitude of PA fluctuation images, whether fluctuations arise from random illuminations or from random distributions of absorbers.
As a first result of our theoretical investigation, we explained why fluctuation PA imaging palliates the visibility artefacts emerging in conventional limited-view and resonant-bandwidth PA imaging.
As a second and major result, we provided with Eq.\ref{eq:fluctuation_image} a general expression of the fluctuation amplitude as a function of statistical properties of the random process inducing fluctuations. 
We consequently demonstrated that fluctuations from red blood cells randomly distributed in blood vessels are expected to be one or two orders or magnitude larger than those expected from multiple speckle illumination of blood vessels at depth in tissue. We recall that this conclusion holds because optical speckle grains at depth are much smaller (half the optical wavelength) than both the acoustic PSF ($h(\rr)$) and the vessel size ($f(\rr))$, as the main assumption of our theory.  These theoretical results quantitatively explain why it has recently been possible to experimentally detect fluctuations from flowing red blood cells (in the context of super-resolution imaging~\cite{chaigne2017super}), whereas the use of speckle-induced fluctuations imaging remain limited to proof-of-concept experiments where speckle grains could be made large enough via free-space propagation. As a general conclusion, our theoretical results suggest that fluctuations from flowing red blood cells at physiological concentrations (around $50\ \%$) may be exploited to remove the visibility artefacts of conventional PA imaging. This has never been demonstrated experimentally so far, and is the objective of the second part of our work.

\section{Experimental results}

\subsection{Phantoms experiments}
As a first experimental demonstration of the proposed approach, we obtained cross-sectional images of the three different types of structure introduced in the simulation part. In each experiment, PA signal fluctuations were produced by a controlled flow of blood at physiological concentration.  

\subsubsection{Experimental setup}
\begin{figure}[htbp]
\centering
\includegraphics[width=\linewidth]{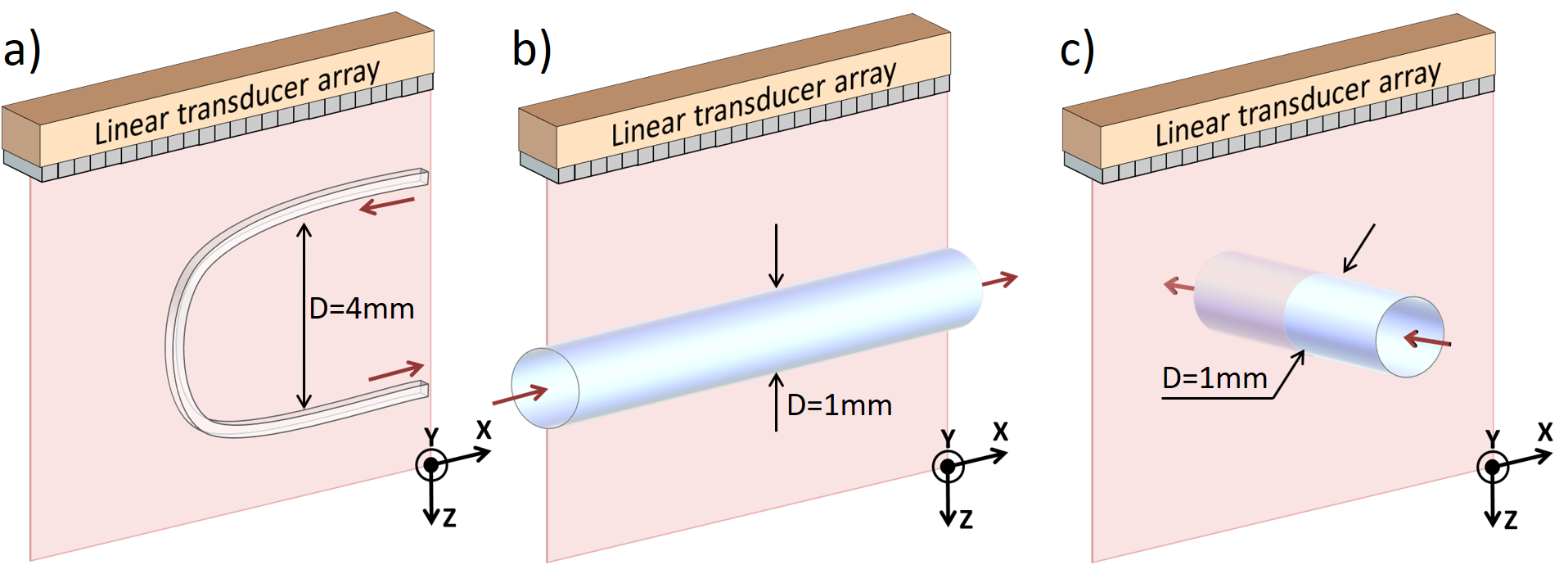}
\caption{Imaging configurations used for the phantoms experiments: (a) in-plane C-shaped capillary, (b) glass tube aligned in-plane , (c) glass tube perpendicular to the imaging plane.}
\label{fig:configurations}
\end{figure}

The three types of experimental configurations are shown in Fig. \ref{fig:configurations}.
The first experiment consisted in imaging a C-shaped structure formed by a polycarbonate (PC) capillary (inner diameter $D$ = 100 $\mu m$, wall thickness $w$ = 20 $\mu m$, Paradigm Optics Inc., Vancouver, USA ) while the second and the third experiments consisted in imaging a cross-section of a glass tube (inner diameter $D$ = 1 mm, wall thickness $w$ = 100 $\mu m$, Capillary Tube Supplies Ltd, UK) aligned within the imaging plane  ($2^{nd}$ experiment) and perpendicularly to the imaging plane  ($3^{rd}$ experiment). Both  samples and the probe were held in a water tank. For each experiment, blood flow was induced through the tubes and capillary via a syringe pump (KDS Legato 100, KD Scientific, Holliston, MA, USA).  For all experiments, the blood velocity was approximately 1 cm/s. 

The samples were illuminated in an oblique manner from the top of the probe, with $\tau$ = 5 ns laser pulses ($\lambda$ = 680 nm, fluence $\approx$ 3 mJ/cm$^2$, pulse repetition frequency (PRF) = 100 Hz) by a frequency-doubled Nd:YAG laser (Spitlight DPSS 250, Innolas Laser GmbH, Krailling, Germany). At each laser shot,  PA signals were recorded with a capacitive micromachined ultrasonic (CMUT) array (L22-8v, Verasonics, USA: 128 elements, pitch $\sim 100 \mu m$, center frequency $f_c\sim 15$ MHz, bandwidth $80\%$) connected to multichannel acquisition electronics (High Frequency Vantage 256, Verasonics, USA).

For each PA acquisition, the raw data was available as a radio-frequency (RF) frame containing the signals received by each transducer element of the CMUT array measured with a sampling frequency $f_s=62.5\ \mathrm{MHz}$. The total number of acquired RF frames was $M_1$ = 1,000 (C-shaped capillary), $M2$ = 1,000 (glass tube, parallel orientation) and $M3$ = 10,000 (glass tube, perpendicular orientation). Given the 100 Hz PRF, the total acquisition times for each experiment was $T_1=T_2$ = 10 s and  $T_3$ = 100 s. A longer acquisition time was chosen for the $3rd$ experiment to ensure that the number of independent red blood cells (RBC) configurations induced by the blood flow was similar in all the three experiments. In contrast to the $1st$ and the $2nd$ experiments, the blood flow in the $3rd$ experiment was oriented perpendicular to the imaging plane. However, the characteristic PSF size in the direction perpendicular to the imaging (xz) plane was about ten times larger (of the order of 1 mm)  than the dimension of the PSF within the imaging plane (of the order of $100\ \mu m$). With the chosen  blood velocity of 1 cm/s, the RBC distribution  over $100\ \mu m$ (respectively 1 mm) distance is typically renewed every 10 ms (respectively 100 ms). Consequently, each of $T_1$, $T_2$ and $T_3$ corresponds to 1000 independent configurations during the measurement time.   

\subsubsection{Data processing}

For each of the three configurations, $N=1000$ independent frames over the total acquisition time were used for data processing. For each experiment, $N=1000$ complex-valued RF frames were computed from the $N=1000$ real-valued frames via a Hilbert transform along the time axis. Complex-valued PA reconstructions $A_k(x,z)$ (with k=1...N) were computed by applying standard delay-and-sum beamforming reconstruction. The mean reconstruction was computed as  $E[A](x,z)=\frac{1}{N}\sum_{k=1}^{M}A_k(x,z)$, and the mean image was defined as  $|E[A](x,z)|$. While the fluctuations of interest were those induced by the blood flow, other sources of fluctuations were present in the experiments, including laser pulse energy fluctuations and detection noise. Because laser pulse energy fluctuations affect identically all PA sources, they may be compensated by pulse-to-pulse energy monitoring. However, this requires  additional hardware and signal processing. 

Here, we used spatio-temporal singular value decomposition (SVD) as a key processing step to filter out laser pulse energy fluctuations (PEF). The SVD approach was first introduced in ultrasound imaging to discriminate tissue and blood motion~\cite{demene2015spatiotemporal} and was used by our group~\cite{chaigne2017super} later on to extract relevant fluctuations in fluctuation-based super-resolution PA imaging. In brief, SVD decomposes the initial data into a basis of spatiotemporal singular vectors. By choosing the singular vectors corresponding to relevant fluctuations, one can suppress signals with different spatiotemporal behaviour such as tissue motion, laser and electronics noise, etc. In our experiments, the first singular vectors (with the highest singular values) corresponded to laser PEF. In the Supplementary Information, we illustrate on both simulation and experimental results how filtering out the first singular values indeed provides  a very efficient way to remove the effect of parasitic laser PEF (see Supplementary information, section 4.A, Fig. S2). We also illustrate the effect of SVD filtering as a function of the range of selected singular values (see Supplementary information, section 4C, Fig. S5). For all experimental images, the captions indicate the range of singular values that were selected (noted SVD cutoffs from now on). 

For each set of N = 1000 SVD-filtered PA images, $A^{\mathrm{SVD}}_k(x,z)$, the standard deviation image was computed through its variance as :

\begin{equation}
\label{eq:sigmasvd}
    \sigma[A](x,z)=\sqrt{\frac{1}{N-1}\sum_{k=1}^{N}\Bigl\lvert A^{\mathrm{SVD}}_k(x,z)-E[A^{\mathrm{SVD}}_k](x,z)\Bigr\rvert ^2}
\end{equation}
We note that the expression above is similar to the one used in ultrasound power Doppler (in which particular case the mean value of the SVD filtered images is zero)~\cite{demene2015spatiotemporal}.

\begin{figure}[htbp]
\centering
\includegraphics[width=\linewidth]{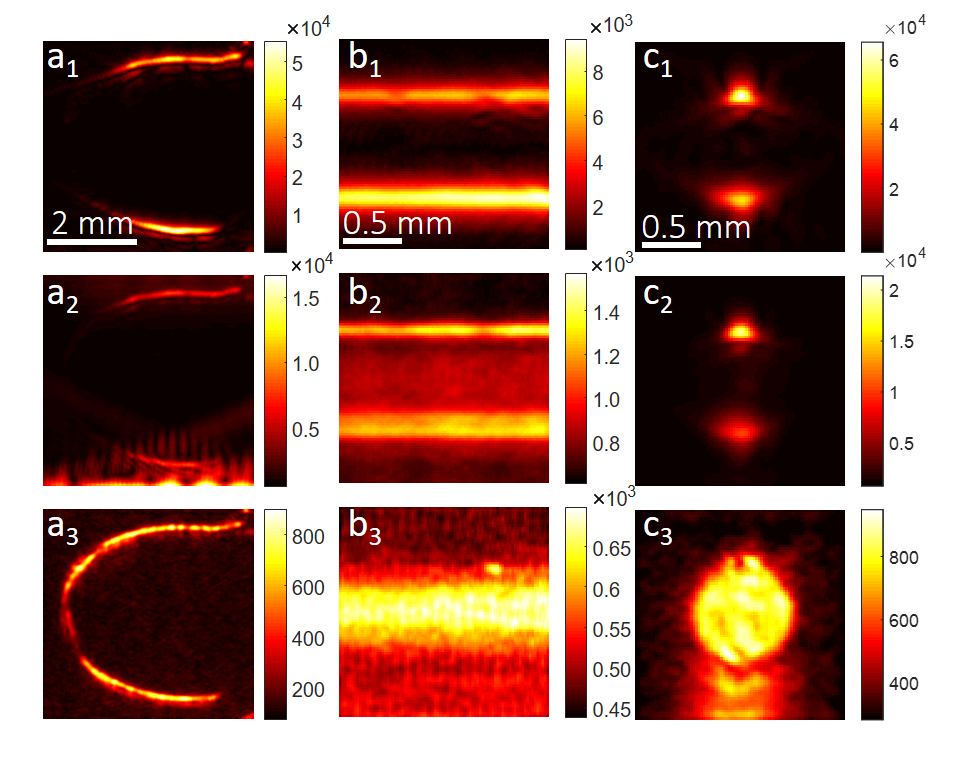}
\caption{Experimental results for the three configurations (a,b and c) introduced in Figs. \ref{fig:SimulationResults} and \ref{fig:configurations}. Row 1: Conventional photoacoustic images (mean). Row 2: fluctuation images (standard deviation) without svd filtering. Row 3: fluctuation images (standard deviation) after svd filtering. SVD cutoffs: a3 [31-61], b3 [11-381] and c3 [15 101].  Fluctuations were induced by blood at physiological concentrations flowing in the structures with a velocity around 1 cm/s.  All images were obtained from processing $N = 1000$ complex-valued RF frames.}
\label{fig:experiments_loop}
\end{figure}

\subsubsection{Results}
\label{subsubsection_Results_phantoms}
The experimental results  are shown in Fig.~\ref{fig:experiments_loop}. Row 1 shows the conventional photoacoustic image. Row 2 shows the fluctuation images that are obtained without the SVD step. Row 3 shows the fluctuation images obtained after SVD filtering.  
These results demonstrate that fluctuations induced by a blood flow at physiological concentration can be exploited to suppress visibility artefacts of conventional PA imaging. Both limited-view and resonant-bandwidth artefacts are indeed well suppressed, in agreement with the simulation results  shown in Fig.\ref{fig:SimulationResults}. The residual signal (clutter) observed at the bottom of the images $b_2$ and $c_2$ is likely due to acoustic reverberation induced by the tube made of glass. This artefact is thus related to the sample rather than the proposed method. The slight inhomogeneity of the fluctuation amplitude within the C-shaped capillary (image $a_2$) may be caused by aberration induced by the capillary and/or an imperfect alignment of the structure in the imaging plane. 

As confirmed by simulations (see Fig. S2 of the supplementary material), the laser pulse energy fluctuations (PEF, measured with a powermeter around $3\%$ relative rms value) completely overwhelms the fluctuation image if not compensated for. Our results illustrate that removing the first singular values efficiently filters out the effect of PEF, and avoid demanding hardware-based monitoring and compensation of those fluctuations. Fig. S5 of the supplementary material further illustrates that once the low singular values are removed to filter out the PEF, the fluctuation images very marginally depends on the choice of SVD upper cutoffs

For the results presented on Fig.\ref{fig:experiments_loop}, N = 1000 frames were used to compute the fluctuation images. In the supplementary information, we present phantoms experimental results obtained from different values of N (Fig. S3, bottom row), which show that depending on the desired image quality, N could be reduced from a a few hundred to a few tens of images, increasing the temporal resolution by the same factor. In addition, the 1D profiles through the fluctuation image presented in Fig. S3 also clearly illustrate that the fluctuation of interest appears on top of a noise fluctuation background caused by detection noise. On the principle, provided that N is sufficiently large, any fluctuation of interest can be detected on top of the noise background, thanks to the additivity of the variance of independent noise sources. Importantly, we note that quantitative measurements (not considered here) requires that the noise background can be estimated and substracted from the variance image. In all the results presented in this paper, the colormap was set such that black corresponded to the background noise on the fluctuation image. 

The dependence on N illustrate the trade-off between acquisition time/temporal resolution and quality of the fluctuation image, which in practice will depend on the level of fluctuation of interest as compared to noise. For the phantom experiments shown here, the peak to noise ratio in the signal domain was around 25-30, but the fluctuation of interest (from blood motion) were about half the noise rms value, which corresponded to a quite realistic situation. Thanks to both the beamforming step (which averages out RF noise by summation over the 128 channels) and the statistical estimation of the fluctuation along N realisations, the fluctuation of interest can in the end be detected in the image domain on top of the noise background even if weaker.

\subsection{\textit{In vivo} experiments}

The proposed approach is further illustrated with a preliminary demonstration of 3D imaging of a live chicken embryo. The chorioallantoic membrane (CAM) of the chicken embryo is composed of small blood vessels whose role is to capture oxygen through the shell and to supply it to the developing embryo. Vessels are spread on a large surface and can have 3D orientation in the vicinity of the embryo, and, therefore, provide a relevant model to assess the feasibility and performance of the proposed approach.

\subsubsection{Experimental setup and methods}

\begin{figure}[h!]
\centering
\includegraphics[width=\linewidth]{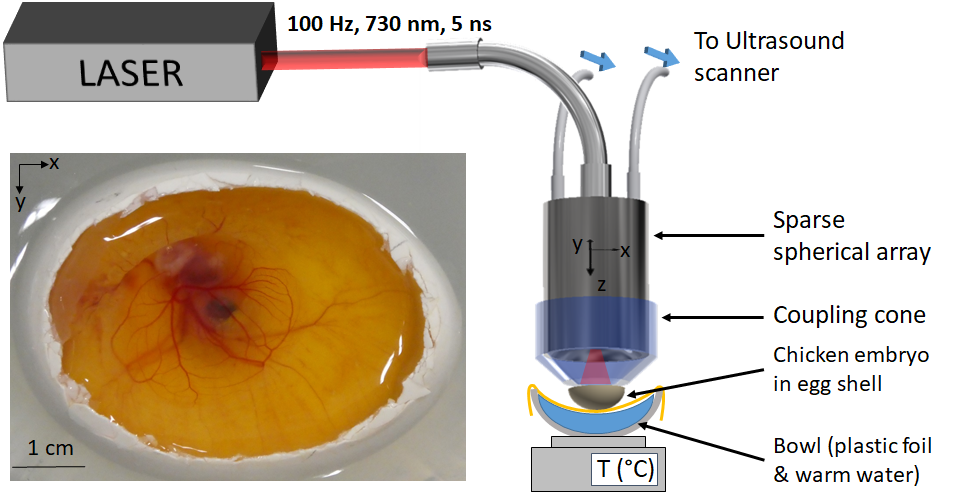}
\caption{Schematic of the experimental setup for 3D \textit{in vivo} imaging of chicken embryo.}
\label{fig:Setup3D}
\end{figure}

The experimental configuration is illustrated on Fig.\ref{fig:Setup3D}. A custom 256-channel spherical transducer array ($8$ MHz, bandwidth $80\%$, $f = 35\ \mathrm{mm}$, $f/D = 0.7$) was designed by our group and fabricated by Imasonics (Voray-sur-L’Oignon, France). This transducer was connected to the same Verasonics acquisition system used in our phantom experiments. An in-house coupling cone, filled with water and closed with a latex membrane, was used to couple the transducer to the sample. Fertilized eggs were obtained from a local farm and placed in an incubator for 6 days. After making a small hole through the shell using a scalpel, 1.5 mL of egg white was removed with a syringe and the top part of the shell was cut with scissors. Warm Phosphate-Buffered Saline (PBS) solution was poured into the shell to ensure acoustic contact with the latex membrane. The sample was held by a hammock made of food wrap. The hammock was attached to a bowl containing water maintained at $36^{\circ}C$ using a hot plate. 
The sample was illuminated with $\lambda$ = $730 nm$ light via a custom fiber bundle (Ceramoptec, Germany) going through the transducer, resulting in a fluence at the top surface of the sample of approximately $3\ \mathrm{mJ/cm^2}$. $M = 768$ frames were acquired at 100 Hz and saved to the computer for off-line data processing. The associated average power density, $300\ \mathrm{mW/cm^2}$, was of the order of the ANSI limit of $230\ \mathrm{mW/cm^2}$. Offline processing was performed exactly as in the 2D phantom experiments, with delay-and-sum beamforming, SVD filtering and statistical computations. 
\subsubsection{{Results}}

\begin{figure}[htbp]
\centering
\includegraphics[width=\linewidth]{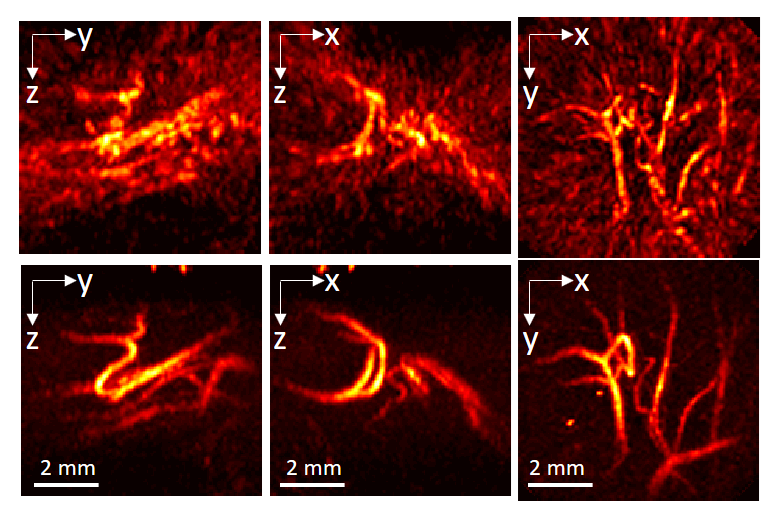}
\caption{\textit{In vivo} images obtained experimentally with the chicken embryo CAM model. Mean intensity projections (MIP) of the 3D images reconstructions along the x, y and z directions. Top row: conventional PA imaging. Bottom row: PA fluctuation imaging. The mean and fluctuation images were obtained from N = 768 frames, corresponding to a measurement time of 7.68s. SVD cutoffs: [25-220].}

\label{fig:experiments_egg}
\end{figure}

The top row of Fig.\ref{fig:experiments_egg} shows the maximum intensity projections (MIP) of the mean image in the three principal directions. These images present strong deterministic background features usually called clutter. In  our experiment, this clutter is likely due to the limited number of transducers ($N$ = $256$) sparsely distributed over the ultrasound probe. Despite the relatively large detection aperture ($f/D$ = $0.7$), several structures oriented predominantly along the z direction (upward and downward vessels) are missing or very weak in this conventional reconstruction, because of the limited-view problem. In contrast, MIPs  of the  3D  fluctuation images (bottom row of Fig.\ref{fig:experiments_egg}) reveal vessels with all possible orientations. This preliminary experiment also suggests that, in addition to removing the visibility artefacts, extracting the flow-induced fluctuations also allows removing the clutter that appears presumably due to the sparse nature of our probe (inducing a 3D PSF with grating lobes).
For the results presented on Fig.\ref{fig:experiments_egg}, all the N = 768 frames (measurement time 7.68 seconds) were used to compute the fluctuation images, with SVD cutoffs = [25-220]. In the supplementary information, we present results obtained from different values of N (Fig. S3) and also illustrate the effect of SVD filtering as a function of singular values cutoffs (Fig. 4C). As observed for the phantoms experiments, filtering out the first singular values is essential to isolate efficiently the fluctuations from blood from those from laser PEF. Fig. S3 shows that N = 128 (measurement time 1.28 seconds) still provides a relatively good image quality, and illustrate the tradeoff between acquisition time/temporal resolution and quality of the fluctuation image. 

The purpose of this preliminary experiment was to illustrate the performance of the method in a biologically-relevant situation, with realistic blood flow. A detailed description and investigation of the performance of our 3D imaging setup are being carried out and was out of the scope of the work presented here. 

\section{Discussion and conclusion}

In conclusion, we demonstrated both theoretically and experimentally that fluctuations induced by blood flow can be exploited to palliate  visibility artefacts of conventional limited-view and resonant-bandwidth PA imaging. We first propose a theoretical framework that provides a quantitative expression of the fluctuation image as a function of the system PSF and  statistical properties of the random medium at stake. Our theoretical framework applies to fluctuations induced either by fluctuating illumination patterns or by random distributions of  absorbing particles (including flowing red blood cells). Our theoretical predictions suggested in particular that  visibility problems could be overcome through exploiting the flow of blood at physiological concentration (overcoming visibility problems had been demonstrated previously only experimentally and those experiments relied on  multiple speckle illuminations under unrealistic conditions). 

We first provided two-dimensional images obtained with vessel-mimicking structures flown with blood at physiological concentration. Finally, we presented a preliminary experiment demonstrating 3D fluctuation imaging \textit{in vivo} of the vasculature of a chicken embryo CAM model. We emphasize that the proposed approach does not require additional hardware and can be implemented on conventional PA equipment since it relies on simple statistical processing. 

The proposed technique has two main potential limitations inherent to its very principle, which are fundamentally related as one may be overcome at the expense of the other : temporal resolution and sensitivity. As discussed in section 3.A.3, and illustrated on Fig. S3 of the Supplementary Information,  any fluctuation of interest can in principle be detected on top of any noise background (sensitivity), provided that the number N of acquisitions is made large enough, at the cost of the temporal resolution.

What we illustrate in the supplementary material (see sections 3 and 4) is that the relevant number to quantify the level of fluctuation is the fluctuation to noise ratio (FNR), rather than the conventional signal to noise ratio on the PA signals. As a rule of thumb, as illustrated by our phantom experiments and additional simulation results provided in Section 4 of the Supplementary Information (Fig. S2), for 128 channels, a FNR on the order of 1 requires $N \sim$ a few hundreds, while a FNR $\sim 10$ requires $N \sim$ a few tens. Importantly, this estimate is valid for N uncorrelated frames, which was the case for our simulations and phantoms experiments. For laser pulse repetition rate high enough, or equivalently for blood flow slow enough, there might be correlations between RF frames, in which case the robustness in the statistical estimation is affected. In particular, it should be kept in mind that samples with different dynamics (such as vascular networks with different vessel sizes and different blood velocities) may be reconstructed with a variable robustness. For instance, an robust reconstruction of blood microcapillaries may require a longer acquisition time that for larger vessels, if the laser repetition period is short enough to resolve the dynamics in the microcapillaries.

In the case of our proof-of-concept \textit{in vivo} experiment, the chicken embryo was mostly transparent,  which is a relatively favorable situation as compared to imaging deep into tissue. The SNR (SNR = peak to noise ratio in the signals domain) in this experiment was on the order of 50, about twice higher than for the phantom experiments. However, the fluence value was approximately $3\ \mathrm{mJ/cm^2}$, whereas this could be increased by a factor ten for deep tissue experiments. In section 6 of the Supplementary material, we predict that the amplitude of the photoacoustic fluctuation image is typically 40 times (for our 15 MHz linear array) to 100 times (for our 8 MHz spherical 2D array) lower than the amplitude of the conventional photoacoustic image, when imaging blood vessels $25\ \mu$m (linear array) and $50\ \mu$m (spherical 2D array). Converted into the signal domain, this predicts that a FNR about 1 requires a SNR of about 20  for our linear array and a SNR of about 50 for our 2D spherical array, which corresponds to the cases of both our phantom experiments and \textit{in vivo} experiments).

Although drawing firm conclusions relative to imaging deep in tissue must involve further \textit{in vivo} investigations in small animal or humans, our results on the chicken embryo suggest that the technique has the potential to be applied for deep tissue imaging, provided that a sufficient fluctuation to noise ratio (FNR) can be reached, i.e. that the number N of acquisitions is sufficiently large. Our estimates above indicate that a SNR around a few tens should provide blood flow a FNR about 1, which should provide very good images for N about a few hundreds.

When compared to deep-learning based approaches to solve the visibility problem, which can be implemented in a single-shot mode, our approach has a lower temporal resolution. It however has the merit of extreme simplicity, and can be implemented straightforwardly, as opposed to deep-learning approaches which requires the availability of a training set, and some quite specific know-how.

The proposed theoretical framework could also be potentially extended to investigate spatio-temporal cross-correlations of fluctuation images, which carry quantitative information on the flow of absorbing particles, and thus may contribute to the field of PA Doppler measurements. Finally, although we focused here on the application to the visibility problem, our theoretical prediction of the second-order fluctuation image will be also be useful to quantitatively exploit fluctuation-based super-resolution imaging.\\

\section{Funding Information}

This project has received funding from the European Research Council (ERC) under the European Union’s Horizon 2020 research and innovation program (grant agreement No 681514-COHERENCE).

\section{Disclosures}

The authors declare no conflicts of interest.

\bibliography{sample}

\bibliographyfullrefs{sample}

\end{document}


\maketitle
\part*{SUPPLEMENTARY THEORETICAL MATERIAL}
\section{Relation between $V_g(\eta)$ and $V_p$}

In the main manuscript, we demonstrated that the variance of PA fluctuations is given by equation (5): 
\begin{equation}
\label{eq:sigma2}
\sigma^2[A](\rr)=\Gamma^2\mu_0^2F_0^2 \sigma_g^2 V_g \cdot [f(\rr)^2 * |h(\rr)|^2]
\end{equation}
where $\sigma_g^2 V_g(\eta) = \int_{\rr} C(\rr)d\rr$ is the volume of the autocovariance function of $g_k(\rr)$. In this supplementary material, we establish the relationship between $V_g(\eta)$ and the particle volume $V_p$, under the assumption that the medium consists of hard identical particles. \\

As a standard tool to describe particles random media, we note $N_{k}(\rr) = \sum_{i} \delta(\rr-\rr_i)$ the random microscopic density which describes the positions of all particles for the $k^{th}$ realization of the random medium. Let $s_p(\rr)$ be the indicator function that defines a single particle (1 inside the particle, 0 elsewhere). The particle volume is given by $V_p = \int s_p(\rr) d\rr$. From these definition, the indicator function $g_k$ of the $k^{th}$ realization of the random medium can be written as:
\begin{equation}
    g_k = N_k\ast s_p
\end{equation}
The autocovariance function of the medium defined as
$$C(\rr_1,\rr_2)=<g_k(\rr_1)\times g_k(\rr_2)>_k-\eta^2$$ may thus be expressed as a function of the autocovariance of the random microscopic density as:
\begin{equation}
    C_g(\rr_1,\rr_2)=<[N_k\ast s_p](\rr_1)\times[N_k\ast s_p](\rr_2)>_k-\eta^2
\end{equation}
Introducing $A_N(\rr_1,\rr_2)=<N_k(\rr_1)\times N_k(\rr_2)>_k$ and $u_p=s_p\star s_p$ as the autocorrelation function of $s_p(r)$, under the assumption of stationarity ($C_g(\rr_1,\rr_2)=C_g(\rr_1-\rr_2)$ and $A_N(\rr_1,\rr_2)=A_N(\rr_1-\rr_2)$), tedious but basic integral calculations lead to the following relationship:
\begin{equation}
    C_g=A_N\ast u_p-\eta^2
\end{equation}
By integrating the equation above, and by use of $\int A_n\ast u_p\dd\rr=\int A_n(\rr)\dd\rr\times \int u_p(\rr)\dd\rr=\int A_n(\rr) V_p^2\dd\rr$, one gets the following expression for $V_g$:
\begin{equation}
\label{eq:V_g_A_N}
    V_g=\frac{1}{\sigma_g^2}\int [A_N\times V_p^2-\eta^2]\dd\rr
\end{equation}
The pair correlation function $B(\rr)$ widely used in condensed matter physics~\cite{Torquato}, also introduced in the context of ultrasound physics\cite{twersky1978acoustic,shung1982ultrasound,Twersky1987low,twersky1991,bascom1995fractal,savery2001point}, is related to the random microscopic density by the following expression~\cite{savery2001point}:
\begin{equation}
\label{eq:A_N}
    A_N(\rr)=m^2B(\rr)+m\delta(\rr)= \left (\frac{\eta}{V_p}\right )^2B(\rr)+\frac{\eta}{V_p}\delta(\rr)
\end{equation}
where $m=\left (\frac{\eta}{V_p}\right )$ is the average concentration of particles.
By combining Eqs.~\ref{eq:V_g_A_N} and \ref{eq:A_N}, and noting that $\sigma_g^2=\eta\times(1-\eta)$, one finally gets the following expression for $V_g$:
\begin{equation}
V_g= V_p\times\frac{1}{1-\eta}[1+\frac{\eta}{V_p}\int (B(\rr)-1) d\rr)]
\end{equation}
$W(\eta)=[1+\frac{\eta}{V_p}\int (B(\rr)-1) d\rr)]$ corresponds to the value for $\mathbf{q}=\mathbf{0}$ of the so-called structure factor $$S(\mathbf{q})=1+\frac{\eta}{V_p}\int[B(\mathbf{r})-1]e^{-i\mathbf{q}\cdot\rr}\dd\rr$$ and is known in ultrasound backscattering theory as the packing factor\cite{twersky1978acoustic,shung1982ultrasound,Twersky1987low,twersky1991,bascom1995fractal,savery2001point}. Finally, one gets the following simple relationship between $V_g$, $V_p$, $\eta$ and $W(\eta)$:
\begin{equation}
\label{eq:VgW}
V_g(\eta)= \frac{W(\eta)}{1-\eta}V_p
\end{equation}

\section{Comparison to ultrasonic backscattering}
From the above expression of $V_g$, the amplitude of the photoacoustic variance image can be expressed as 
\begin{equation}
\label{eq:varPA}
\sigma^2[A](\rr)=\Gamma^2\mu_0^2F_0^2 \eta V_p W(\eta) \cdot [f(\rr)^2 * |h(\rr)|^2]
\end{equation}
We note that the dependence of the photoacoustic variance as a function of $\eta$ and $W(\eta)$ is identical to that of the ultrasound backscattering coefficient BSC reported in the field of ultrasound imaging of blood as~\cite{bascom1995fractal}:
\begin{equation}
\label{eq:BSC}
BSC = \mu_s \eta W(\eta)
\end{equation}
where $\mu_s=\sigma_s \frac{\eta}{V_p}$ is the ultrasound scattering coefficent ($\sigma_s$ being the scattering cross-section of a single particle).\\

It was out of the scope of our experimental work here to confront our theory with respect to its dependency on the volume fraction. However, it was recently reported from photoacoustic experiments on blood flow that the dependence of the amplitude of photoacoustic fluctuations (expressed as variance) as a function of blood concentration agreed with the dependence predicted from the backscattering theory as $\eta W(\eta)$~\cite{bucking2019acoustic}, which thus support our theoretical results. In these experiments, the best agreement was obtained with the expression of the packing factor of hard spheres (m=3), given by~\cite{bascom1995fractal}:
\begin{equation}
W(\eta)= \frac{(1-\eta)^{m+1}}{(1+(m-1)\times\eta)^{m-1}}
\end{equation}
With $\eta\sim50\ \%$ (typical volume fraction for blood), the formula above gives $W\sim 16\ \%$, which is the value that was used in the main text to compare the fluctuations from blood to that from multiple speckle experiments.

A detailed comparison of Eqs.~\ref{eq:varPA}  and \ref{eq:BSC} is out of purpose at this stage, given the quite different conditions for which these expressions have been obtained: Eq. \ref{eq:BSC} for the backscattering coefficient is obtained under the assumption of harmonic ultrasound, and do not take into account any specific detection geometry or resolution, while 
Eq.~\ref{eq:varPA} is a theoretical prediction of a variance \textit{image}, including both the properties of the pulsed imaging system (through its PSF) and the statistical properties of the medium.
\clearpage

\part*{FLUCTUATIONS OF INTEREST vs PARASITIC FLUCTUATIONS}

In this second part of the Supplementary information, we investigate and discuss the influence of parasitic sources of fluctuation on the estimation of the fluctuation of interest. In photoacoustic fluctuation imaging, the two main sources of parasitic fluctuations are the electronic noise present on the measured RF signals, and the shot-to-shot energy fluctuation of the laser source. Temporal jitter between the ultrasound acquisition and laser pulses may also be an additional source of fluctuation, depending on the level of jitter as compared to the ultrasound time scale. In our experiments, the temporal jitter was on the order of 1 ns, and its effect was negligible as compared to that of the shot-to-shot energy fluctuations given our ultrasound frequency range, as verified through simulations. We thus limit our discussion here to the effects of the electronic noise and shot-to-shot laser energy fluctuations.

We first introduce the relevant quantities used to measure the various levels of fluctuations,  in both the signal and image domains. We then investigate the effects of the electronic noise and shot-to-shot laser energy fluctuations on the performances of photoacoustic fluctuation imaging. In particular, we investigate the influence of the number N of images to acquire on the quality of the fluctuation image, on both simulations and experimental results. We also explain and illustrate in some details the effect of the SVD processing step, including a sensitivity analysis to the choice of SVD filtering cutoffs. Finally, based on the theoretical expressions provided in the main manuscript, we predict and compare quantitatively the amplitude of the conventional photoacoustic signals/images to that of the fluctuation signals/images, in the context of imaging small blood vessels (with both our linear and 2D arrays).\\

\section{Definitions of relevant noise-related quantities}

\begin{figure*}[!h]
\centering
\includegraphics[width=1\linewidth]{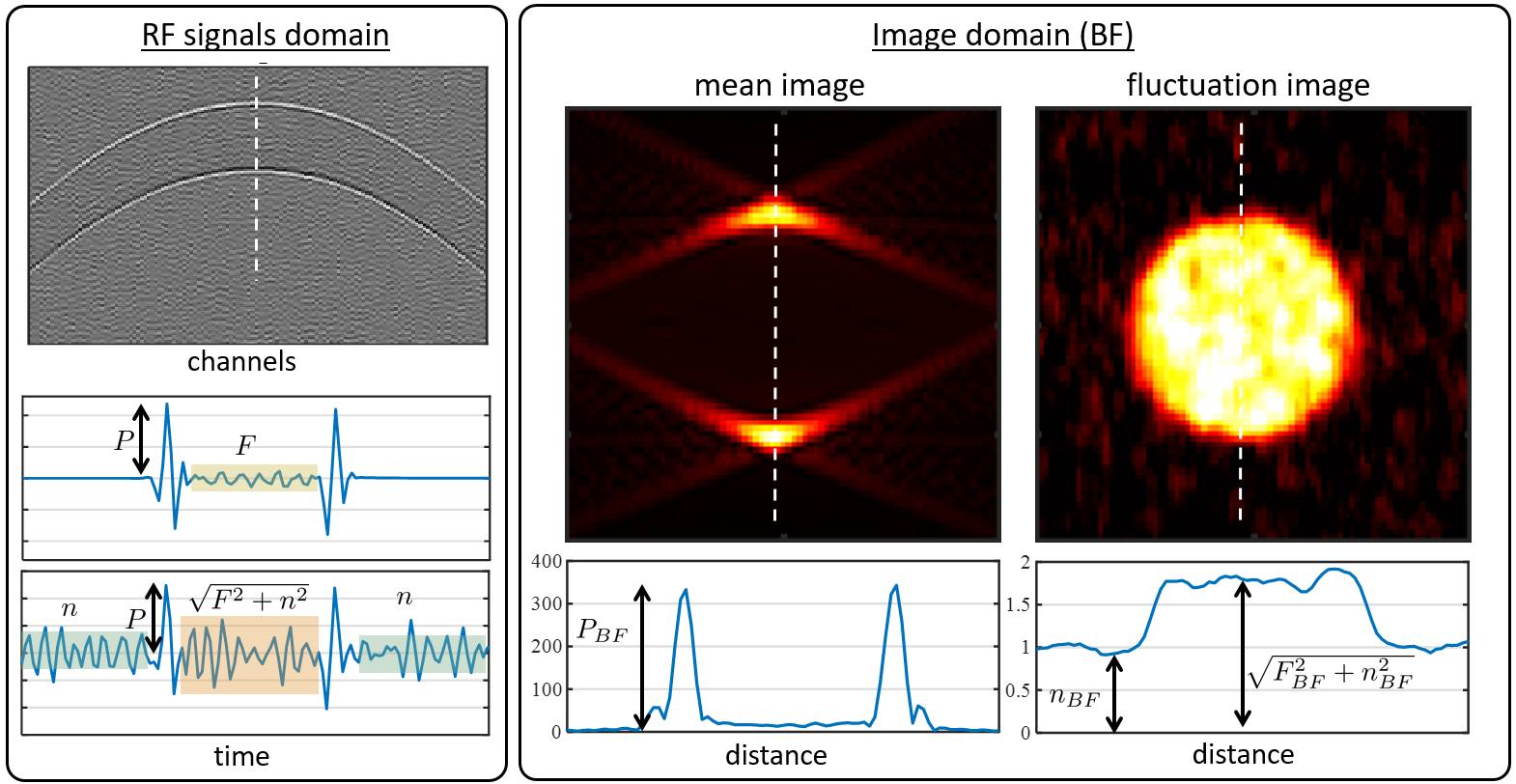}
\caption{Definition of noise and fluctuation levels in signals domain (RF) and image domain (BF). The simulation results corresponds in this example to PNR=5, FNR=0.5. The temporal and spatial profile are taken along the vertical dashed lines.}
\label{fig:NoiseDefs}. 
\end{figure*}
In Fig. \ref{fig:NoiseDefs}, we define the relevant noise, fluctuation and peak amplitude, in both the signals domain and the image domain. We used the fact that the sum of independent random variables has a total variance given by the sum of each variance. In conventional ultrasound signal processing, the SNR would be defined as P/n. In the context of fluctuation imaging, the useful "signal" is rather a fluctuation signal, noted F. To avoid confusion, we defined the peak-to-noise ration PNR = P/n, the fluctuation to noise ratio FNR = F/n. As illustrated in the image domain, the relevant "SNR" quantity for fluctuation imaging is given by $$FNR^{BF}=\frac{F_{BF}}{n_{BF}}$$
While the value of $FNR^{BF}$ in the image domain is similar to the value of $FNR$ in the signals domain, they are strictly speaking not identical as the noise is usually uncorrelated between channels, whereas there may be some spatial coherence in the fluctuation from real sources. As independent fluctuations are summed incoherently, whereas PA signals are summed coherently during the beamforming step, the peak-to-fluctuation ratio PFR in the signals domain is usually extremely different from $$PFR^{BF}=\frac{P_{BF}}{F_{BF}}$$ defined from the comparison between the mean image and the fluctuation image. The possibility to estimate robustly a fluctuation image depends fundamentally on the $FNR$ (i.e. on the detection noise, laser pulse energy, transducer sensitivity, etc.) while the $PFR$ only depends on the absorbers distribution and the system PSF.

\section{Fluctuation of interest vs parasitic fluctuations}

\subsection{removing pulse energy fluctuations with SVD filtering.}

\begin{figure*}[!h]
\centering
\includegraphics[width=\linewidth]{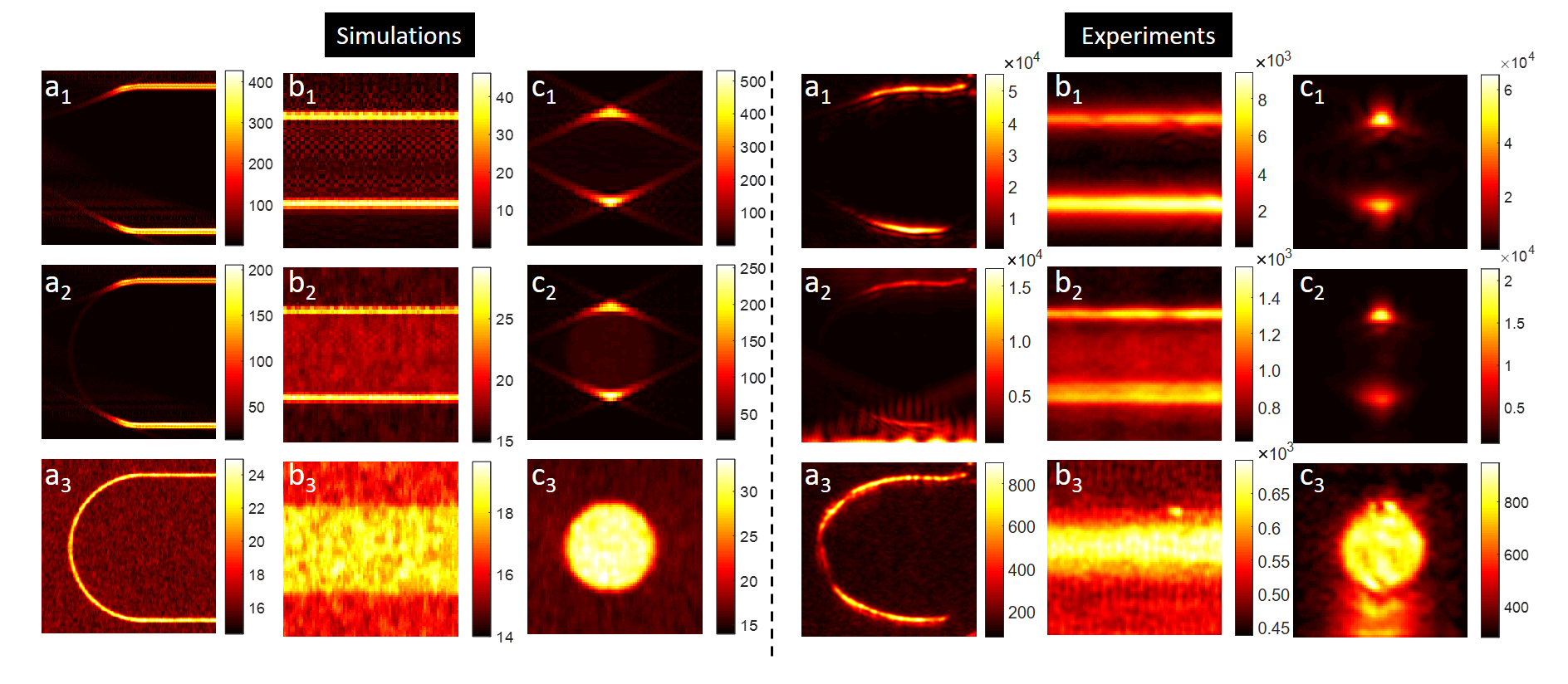}
\caption{Illustration of PEF removal with SVD filtering. Row 1: mean image. Row 2: fluctuation images obtained  without SVD filtering. Row 3: fluctuation images after SVD filtering. Pulse energy fluctuations were present on both simulated and experimental signals.}
\label{fig:SVD_PEF}. 
\end{figure*}
In Fig. \ref{fig:SVD_PEF}, left panel, we provide simulations images that were obtained in extremely noisy conditions: SNR = 5, FNR = 0.5 (RF noise twice larger than the fluctuations of interest), and relative pulse energy fluctuations of $50\%$. The second raw illustrate the results of computing the fluctuation image without prior SVD filtering. In this condition, the fluctuation is completely dictated by the PEF, and what is seen is the average photoacoustic source distribution that fluctuates with the PEF. When just the first singular value is removed, the effect of the PEF is completely filtered out, and is image is visually identical to that observed without PEF in the simulation. The same effect is observed for the phantom experiment on all three types of structures (as already shown in Fig.5 of the main manuscript). The only difference between simulations and experiments is that it is necessary to remove more than one singular value to filter out the PEF in the experimental situations. We have no clear explanation for this at that stage. The sensitivity to the choice of SVD filtering cutoffs is illustrated further below (Fig.\ref{fig:FigS5_svd_sensitivity}).

\subsection{Fluctuation of interest vs electronic noise}

\begin{figure*}[!h]
\centering
\includegraphics[width=\linewidth]{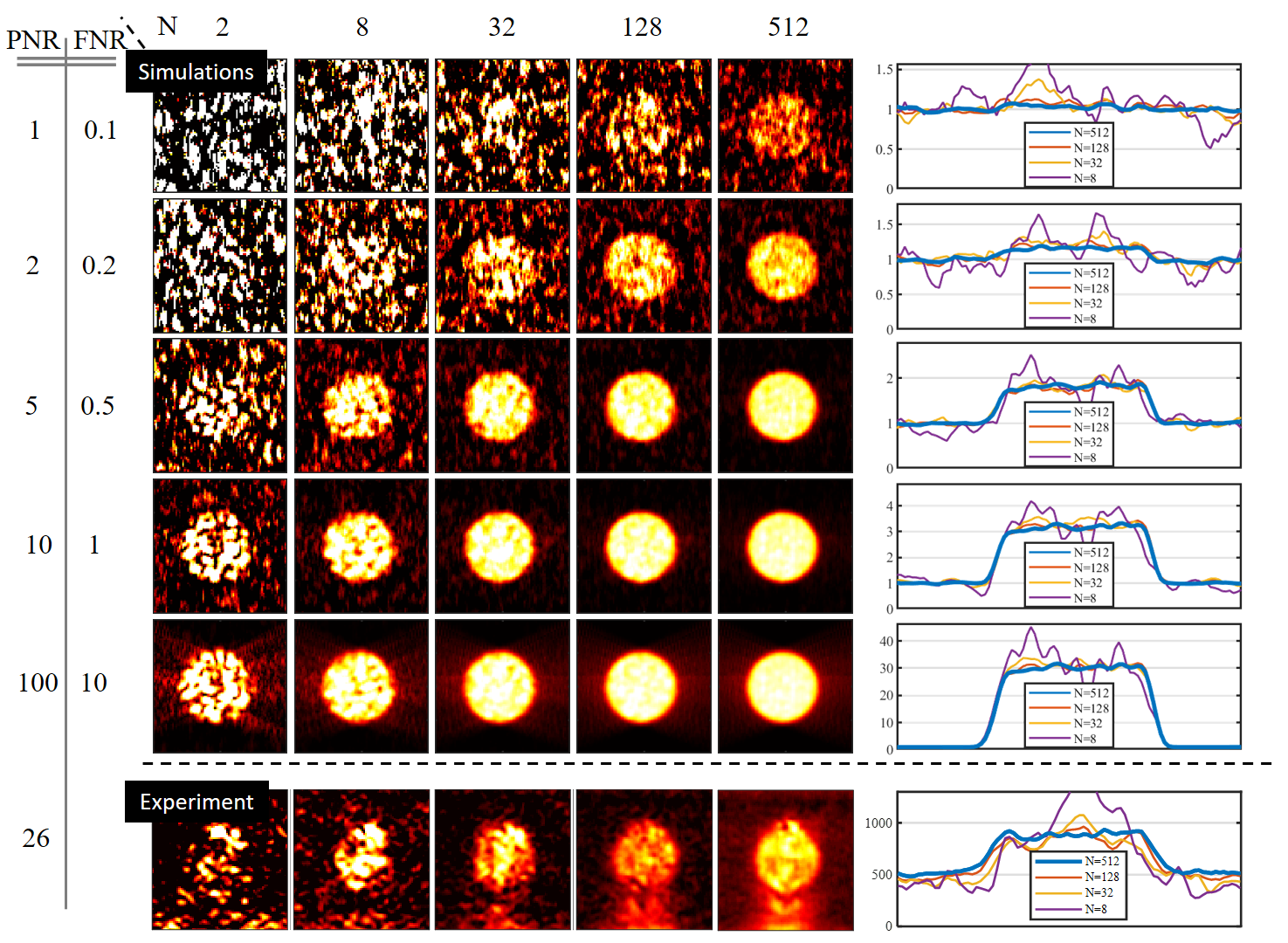}
\caption{Illustration of the effect of the electronic noise and the number of measurement N on the fluctuation image. The PNR was fixed for the experiment (bottom row). 1D profiles through the fluctuation images are shown on the last column. The colormap for each image was adjusted such that black corresponds to the noise floor, and yellow corresponds to the average level in the object.} 
\label{fig:SNR_simu_expDisk}. 
\end{figure*}

\begin{figure}[!h]
\centering
\includegraphics[width=0.7\linewidth]{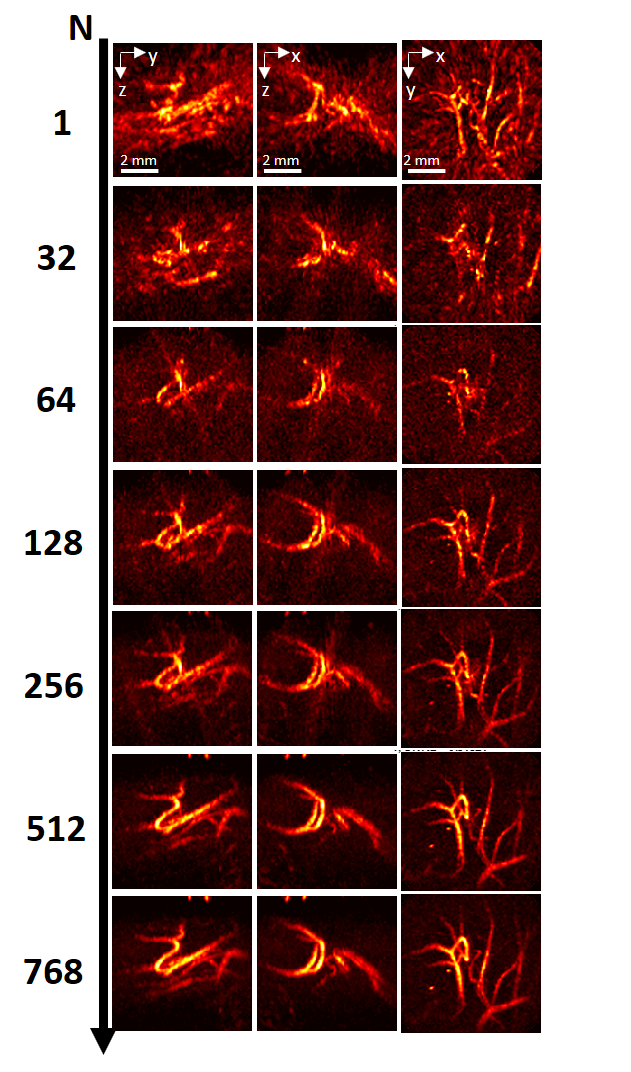}
\caption{Evolution of the image quality as a function of N for the chicken embryo experiment. Each volume is represented as 2D MIP (max intensity projection) image (yz, xz and xy planes).}
\label{fig:Embryo_N}. 
\end{figure}

Fig.\ref{fig:SNR_simu_expDisk} illustrates both the effects of the electronic noise and of the number N of acquisition. As a key  conclusion, the 1D profiles through the fluctuation images shows that the object appears on top of the noise background. The level of noise is actually only dictated by the electronic noise (value 1 in the image space, as our choice of unit). The ability to detect the object on top of the noise is actually depending on the statistical power with which both the noise and the fluctuation of interest are estimated: for large FNR values, the noise is negligible, and the only fluctuation comes from the fact that N is finite. When noise is significant (FNR on the order of 1 or lower), it is always possible to detect the additional fluctuation from the object, but at the cost of a large N. For the phantom results shown at the bottom row (corresponding to the experiment used for Fig. 5), the SNR was about 25, and the FNR about 0.8. Fig.\ref{fig:SNR_simu_expDisk} clearly demonstrates that the choice of N is dictated by the FNR value, and a trade-off between temporal resolution and robust estimation of the object. Fig.\ref{fig:Embryo_N} illustrates this trade-off for the chicken embryo experiments.\\

\clearpage

\subsection{SVD filtering: sensitivity to upper and lower singular values}

\begin{figure*}[!h]
\centering
\includegraphics[width=\linewidth]{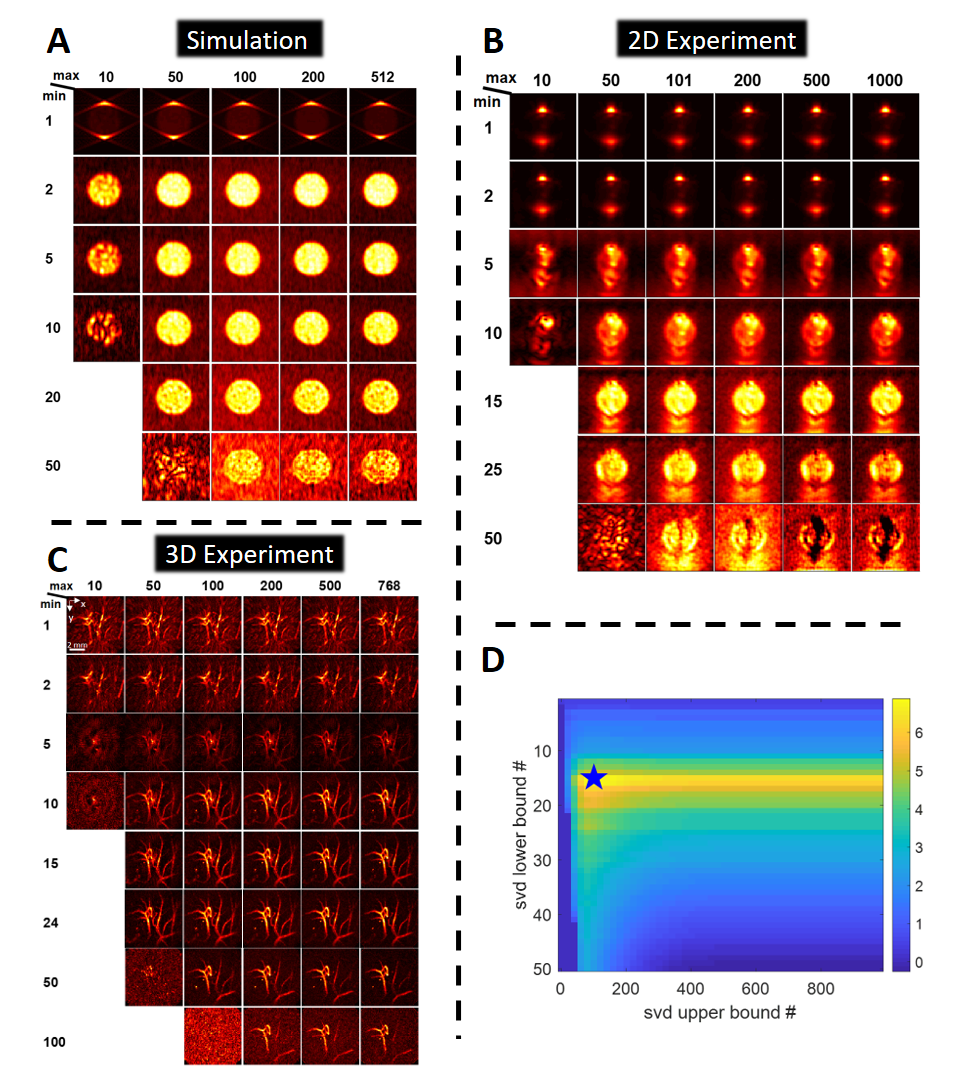}
\caption{SVD sensitivity analysis: A-C Fluctuation images obtained  for several svd cutoff [min-max]. A: Disk simulation, B: Phantom experiment, C: chicken embryo experiment. D: Contrast-to-Noise ratio (CNR) map for phantom experiment.}
\label{fig:FigS5_svd_sensitivity}. 
\end{figure*}

Fig.\ref{fig:FigS5_svd_sensitivity} provides a sensitivity analysis to the choice of the SVD filter cutoffs. The main message from this sensitivity plot is that removing  the first singular values is essential to filter out the effect of laser PEF. While removing just one the first singular value allowed to perfectly removed the effect of the PEF in simulations, it is necessary to remove more for the experiments. We have no clear explanation for this at that stage. One other important conclusion is that filtering out high singular values has little effect on the final image. We observe on quantitative analysis of the fluctuation images that both the noise and the object are decreased when high singular values are filtered out. We observed that depending on the image quality criterion, filtering some high singular values can slightly improve the image quality. This explain why we present images with both low and high singular values cutoff. Fig.\ref{fig:FigS5_svd_sensitivity}D, obtained for the phantom experiment (B), illustrates how a CNR metrics (defined as $CNR = \frac{\text{Fluctuation in object - background fluctuation level}}{spatial variance of ensemble variance in background + spatial variance of the ensemble variance in object}$) is affected by the choice of SVD cutoffs. This figure illustrate that from a CNR point of view, the key choice is that of the low SVD cutoffs: too low would leave some PEF, while too high degrades the object of interest. This metrics is very marginally affected by the choice of the high SVD cutoff.

\section{$\text{PFR}^\text{BF}$ prediction for vessel-like structure}

In this section, we use equations (4) and (5) of the main manuscript to numerically compute the ration $R(\rr)$ between the mean photoacoustic image and the fluctuation image, for vessel like structures with various orientation, in 3D.

\begin{equation}
\label{eq:MeanVSfluc}
    R(\rr)=\frac{E[A](\rr)}{\sigma[A](\rr)} = \sqrt{\frac{\eta}{W(\eta) V_p}} \frac{f(\rr) \ast h(\rr)}{\sqrt{f^2(\rr) \ast |h|^2(\rr)}}
\end{equation}

\begin{figure}[!h]
\centering
\includegraphics[width=0.5\linewidth]{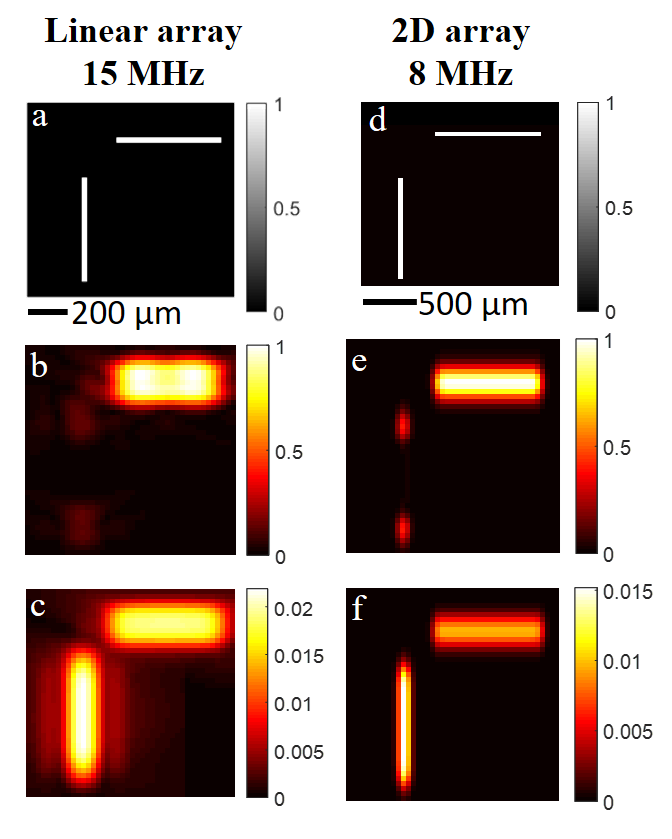}
\caption{Computation of the theoretical mean and fluctuation images for 2D imaging with a linear array (left column) and for 3D imaging using a 2D array (right column). a,d Object f(r) in the xz plane. b,e Mean conventional PA image. c,f Fluctuation image. Images b-c and e-f are normalized with the corresponding maximum values of conventional PA image.}
\label{fig:TheoPSF}. 
\end{figure}

This ratio allows a quantitative comparison between the two modalities in given situations defined by a 3D object $f(\rr)$ and the 3D PSF, and corresponds to a local value (computed pixel wise) of the $PFR_{BF}$ as defined in section 3. A blood flow is considered with $\eta = 50 \%$, $W(50\%) = 16 \%$ and $V_p = 100 \mu m^3$. The signals to reconstruct the 3D PSF were computed for our linear array (L22-8v) using Field-2 \cite{Jensen96field:a}\cite{jensen139123} and 
measured for our custom 2D array. We computed (\ref{eq:MeanVSfluc}) for the case of a horizontal and a vertical thin vessel ($\lambda/4$ diameter, $\lambda/4=24 \mu m$ for the linear array @15MHz and $\lambda/4=47 \mu m$ for the 2D array @8 MHz) (see Fig.\ref{fig:TheoPSF}a-d). With this object geometry, conventional mean PA image suffers from visibility artefacts only for the vertical vessel as shown in Fig.\ref{fig:TheoPSF}b-e. The fluctuation images show no visibility artefact (Fig.\ref{fig:TheoPSF}c-f). Images are normalized by the maximum of the conventional mean images. 

R values are evaluated by taking the ratio of the images at the center of the object $\rr_{obj}$ for the different types of vessel, horizonal and vertical.
Table \ref{tab:R} indicates the computed values of R. As an indication, we also computed R for the case of a transverse (Transv.) vessel for the 2D imaging case, ie. going across the imaging plane. The ratio between the fluctuation at any configuation $r_{obj}$ and the value of $E[A](\rr_{hori})$ in the absence of visibility artefact is also computed at the bottom line.

\begin{table}[htbp]
\centering
\caption{$\bf R(\rr_{obj})={E[A](\rr_{obj})}$/${\sigma[A](\rr_{obj})}$}
\begin{tabular}{*6c}
\toprule
 {} &  \multicolumn{3}{c}{2D 15 MHz} & \multicolumn{2}{c}{3D 8 MHz}\\
\midrule
\hline
 {} & Horiz. & Vert. & Transv. & Horiz. & Vert. \\
\hline
$R(\rr_{obj})$   & 43.3 & 0.032 & 94.6 & 103.1 & 0.38 \\
$\frac{E[A](\rr_{hori})}{\sigma[A](\rr_{obj})}$ & 43.3 & 38.8  & N.A & 103.1 & 65.8 \\
\hline
\bottomrule
\end{tabular}
  \label{tab:R}
\end{table}

The fluctuation is about 40x smaller than the mean PA image in the in-plane 2D imaging case (vessel diameter $\lambda/4$, 15 MHz). In the transverse flow case, the anisotropy of the PSF induces a different value around 100. In 3D imaging, the PSF is isotropic and the values for both the horizontal and vertical vessels are close to 100. It should be kept in mind that these values vary with the frequency/geometry of the probe and the considered vessel size. 
However, the quantitative evaluation of $R(\rr_{obj})$ offers a way to predict the fundamental limits in the detection of PA fluctuations at depths in the image space, as this ratio is independent of the fluence. 

\clearpage

\bibliography{sample}

\bibliographyfullrefs{sample}